\documentclass[
preprint,
 amsmath,amssymb,
 aps,
]{revtex4-1}
\usepackage{eqnarray,amsmath}
\usepackage{graphicx}
\usepackage{dcolumn}
\usepackage{bm}
\usepackage{hyperref}
\usepackage{color}
\usepackage{tabularx}
\usepackage{multirow}

\begin{document}


\title{Effect of oblique horizontal magnetic field on convection rolls}

\author{Snehashish Sarkar$^{1}$, Sutapa Mandal$^{2}$, and Pinaki Pal$^{1}$}
\affiliation{$^{1}$Department of Mathematics, National Institute of Technology, Durgapur~713209, India\\
$^{2}$School for Engineering of Matter, Transport and Energy, Arizona State University, Tempe, AZ 85287, USA}


\begin{abstract}
We investigate the effect of external horizontal magnetic field applied on the convection rolls obliquely (at an angle $\phi$ with the $x$-axis) in electrically conducting low Prandtl number fluids under the paradigm of the Rayleigh-B\'{e}nard convection by performing three-dimensional direct numerical simulations. The control parameters, namely, the Chandrasekhar number ($\mathrm{Q}$) and the reduced Rayleigh number $r$ (ratio of Rayleigh number to critical Rayleigh number), are varied in the ranges $0 \leq \mathrm{Q} \leq 1000$ and $1 \leq r \leq 20$ for the Prandtl numbers $\mathrm{Pr} = 0.1$ and $0.2$ by considering three horizontal aspect ratios ($\Gamma$): $\frac{1}{2}$, $1$ and $2$. In the absence of the magnetic field, the convection starts in the form of steady rolls including the one parallel to the $x$-axis. As the oblique horizontal magnetic field is switched on at an angle $\phi \in (0^\circ, ~90^\circ]$ with the $x$-axis, it is observed that the Lorentz force generated by the component of the magnetic field transverse  to the axis of the convection rolls inhibits convection in the form of steady rolls. Thus, with the application of the magnetic field, the convection is suppressed and restarts for a higher Rayleigh number in the form of steady convection rolls. The rolls can either be oriented along the $x$-axis (steady parallel rolls, SPR) or oriented at an angle $45^\circ$ (steady oblique rolls, SOR$^+$) with the $x$-axis depending on the choices of the parameters. 
A rich bifurcation structure consisting of standing and traveling flow patterns associated with these steady flow patterns for higher values $r$ is investigated in detail. The oscillatory instability of the steady rolls is found to scale as $\mathrm{Q}^\alpha$ with two distinct exponents, one each for weaker and stronger magnetic fields. The investigation reveals that for a given set of values of $\mathrm{Q}$ and $\mathrm{Pr}$, the heat transfer is inhibited with the increase of $\phi$.

\end{abstract}

\maketitle

\section{\label{sec:introduction}INTRODUCTION}

Magnetoconvection, the interaction between thermal convection and magnetic fields, is a phenomenon of significant importance across multiple disciplines, ranging from astrophysics to industrial applications~\cite{spiegel:1969,hurle:1972,glatzmaier:1999,davidson_ARFM:1999,traxler:2011a}. This interplay governs the behavior of electrically conducting fluids in natural settings and engineered systems. In astrophysics, magnetoconvection is crucial for understanding stellar dynamics, influencing phenomena like sunspot formation \cite{schrijver:2008} and solar flares \cite{stein:1998}. It also plays a key role in modeling planetary interiors, such as Earth's outer core, where the geodynamo generates the planet's magnetic field \cite{glatzmaier_Roberts:1995}. In nature, this process occurs in the movement of molten metals within planetary cores, affecting magnetic field generation and planetary evolution \cite{stevenson:2003}. Industrially, magnetoconvection is applied in processes like crystal growth, where controlled magnetic fields improve material properties \cite{hurle:book_1994}, and in metallurgy, where they regulate heat and mass transfer to enhance product quality \cite{davidson_ARFM:1999}. A specific and well-studied example of magnetoconvection is Rayleigh-B\'{e}nard magnetoconvection (RBM), where a fluid layer, heated from below and cooled from above, is subjected to an external magnetic field. 

The dynamics of RBM are governed by four dimensionless parameters: the Rayleigh number (Ra), Chandrasekhar number (Q), thermal Prandtl number (Pr), and magnetic Prandtl number (Pm). Over the years, by examining the RBM system researchers developed a deeper understanding of the complex interplay between thermal convection and magnetic field, revealing the fundamental mechanisms that drive many natural as well as industrial processes~\cite{savage:1969,hurle:1972,olson:2011}.

Different aspects of convection including instabilities~\cite{bodenschatz:ARFM_2000,busse:1971,clever:POF_1990,dan:2014}, patterns~\cite{cross:RMP_1993,rebecca:book,pal:PRE_2013}, chaos~\cite{knobloch_JFM:1986,paul:2011,nandu:EPL_2015}, turbulence~\cite{swinney_gollub:book_1985,manneville_book:2010}, heat transfer~\cite{loshe:RMP2009,Olson:2001}, flow reversals~\cite{yanagisawa_PRE:2011,mannattil_EPJB:2017,mandal:2021,sharma:PRF_2022} etc. have been investigated by the researchers using the Rayleigh-B\'{e}nard geometry and revealed many interesting results. Here we mention some of those relevant to the present study. Fauve et al.~\cite{fauve_et.al:JPL_1981} performed one of the early experiments to investigate the effect of longitudinal and transverse horizontal magnetic field on the oscillatory instability (OI) of the convection rolls in mercury~\cite{fauve:1984}. It has been revealed that the OI is suppressed by both the longitudinal and transverse magnetic fields. However, while suppressing the OI, the longitudinal and transverse magnetic fields show opposite effects on the frequency of oscillation of the related periodic solution. In the first case, the frequency is substantially enhanced, while, in the second case it is slowly reduced.  Interestingly, it has been shown both experimentally~\cite{fauve:1984} and numerically~\cite{busse:1983,pal:2012,sharma:PRF_2022,sharma:2023} that the onset of the OI of convection rolls scales as $\mathrm{Q}^{\alpha}$, with $\alpha \sim 1.2$ for low Prandtl number fluids in the presence of longitudinal magnetic field. Close to the onset, the longitudinal magnetic field not only affects the OI but also impacts the route to chaos. Experiment~\cite{fauve_prl:1984} as well as numerical simulations~\cite{nandu:2015} show period doubling and quasiperiodic routes to chaos in low Prandtl number fluids for weaker and stronger longitudinal horizontal magnetic fields respectively. Researchers also explored the transitions in liquid metals and reported a variety of flow regimes including the one characterized by the phenomenon of flow reversals in RBM~\cite{fauve_prl:1984,yanagisawa:2013,yanagisawa_PRE:2015}.

Apart from the aforementioned works, in the presence of the transverse magnetic field, early studies \cite{shercliff:1953, murgatroyd:1953, lock:1955} concentrated on the classic MHD flow geometries, including pipes, channels, and parallel planes. These works primarily explored how magnetic fields stabilize the flow, reducing turbulence and enhancing flow stability. The later studies \cite{priede:2010, priede:2012, priede:2016} focused on duct geometries, particularly examining the effects of thin and conducting walls in the presence of strong magnetic fields on liquid-metal flows. These studies observed the formation of jets along conducting walls and identified the critical conditions for flow instabilities in the presence of a strong external magnetic field. Recently, Sharma et al. \cite{sharma:2023} numerically investigated the effect of transverse magnetic field on the OI, and the transition scenario near the onset of electrically conducting low Prandtl number fluids for different aspect ratios.

In all the works mentioned so far, the external magnetic field is either applied in the vertical or horizontal (longitudinal or transverse) directions. However, in many practical situations, such as the Earth's core, solar dynamics, and various industrial processes, magnetic fields are often not perfectly aligned with the vertical axis \cite{glatzmaier:1995, miesch:2005}. In applications like fusion reactors or metallurgical processes, oblique magnetic fields can be utilized to stabilize or manipulate fluid flows, thereby enhancing efficiency and safety \cite{davidson_ARFM:1999}. These fields introduce new phenomena, including anisotropic convection patterns and traveling wave solutions, which improve our fundamental understanding of magnetohydrodynamics (MHD) and fluid dynamics \cite{proctor:1982}. Notwithstanding, research on oblique magnetic fields remains relatively limited, highlighting the need for further exploration in this area. Most of the works in the presence of oblique magnetic field, it is applied at an angle with the vertical axis. Chandrasekhar \cite{chandra:book} analytically determined the critical wavenumber and Rayleigh number for rigid boundaries in the presence of a magnetic field inclined in the x-z plane. The influence of an oblique magnetic field on solar oscillation frequencies and compressible convection in a stratified atmosphere has also been extensively studied \cite{philip:1992, matthews:1992, hurlburt:1996}. Julien et al. \cite{julien:2000} investigated two-dimensional convection under a strong oblique magnetic field, showing that stronger fields align fluid motion with the magnetic field lines, leading to structured convective patterns and hysteresis during transitions between vertical and horizontal modes. Thompson \cite{thompson:2005} further noted that an inclined magnetic field breaks the system's reflectional symmetry, producing traveling wave solutions rather than steady states, with oblique rolls dominating at higher tilt angles. Naffouti \cite{naffouti:2014} found that increasing the magnetic field strength generally suppresses convection and reduces heat transfer, but maximum heat transfer occurs when the magnetic field is inclined at an angle $\alpha=30^\circ$ with the vertical, while vertical alignment ($\alpha=90^\circ$) results in the poorest heat transfer. 

However, in none of the studies, the effect of an oblique horizontal magnetic field inclined at an angle with the roll axis has been investigated which we consider in this paper. The oblique horizontal field applied on the convection rolls is found to significantly affect the flow of electrically conducting low Prandtl number fluids in a wide region of the parameter space defined by $0 \leq \mathrm{Q} \leq 1000$ and $1 \leq r \leq 20$ for the Prandtl numbers $\mathrm{Pr} = 0.1$ and $0.2$. Several aspects of convection including the bifurcation structure, instabilities, patterns dynamics, heat transfer, etc. are investigated in detail. The investigation not only reveals the inhibitory effect of oblique magnetic field on the convection rolls but also significantly affects the oscillatory instability of the convection rolls exhibiting two distinct scaling laws. Our numerical investigation also reveals the effect of the angle of application of the magnetic field on the heat transfer in the system.  In the following section, we describe the physical system considered for the study.

\section{System Description} \label{sec:system}
In this paper, we consider the classical Rayleigh-B\'{e}nard convection (RBC) setup. The physical system of RBC consists of a thin layer of Boussinesq fluid of thickness $l$, coefficient of volume expansion $\alpha$, kinematic viscosity $\nu$, thermal diffusivity $\kappa$, and magnetic diffusivity $\eta$ kept between two horizontal, thermally and electrically conducting plates. An adverse temperature gradient $\beta = \Delta T/l = (T_l - T_u)/l$ is maintained between the plates by keeping the temperatures  $T_l$ and $T_u$ of the lower and the upper plates respectively fixed with $T_l > T_u$. Here we intend to investigate the effect of uniform oblique horizontal magnetic field  $({\bf B}_0 = (B_0 \cos \phi, B_0 \sin \phi,0))$ on the flow patterns. Thus, in the presence of the external oblique magnetic field, the convective motion of the fluid is governed by the following dimensionless equations

\begin{subequations}
\begin{gather}
\frac{\mathcal{D} {\bf u}}{\mathcal{D} t} - \mathrm{Q} \left[ \cos \phi\frac{\partial {\bf b}}{\partial x} + \sin \phi\frac{\partial {\bf b}}{\partial y} + \mathrm{Pm}({\bf b}\boldsymbol{\cdot \nabla}) {\bf b} \right]  = -\pmb{\nabla} p + \mathrm{Ra}\theta {\bf \hat{k}} + \nabla^2 {\bf u} , \label{ME}\\ 
\mathrm{Pm} \left[\frac{\mathcal{D} {\bf b}}{\mathcal{D} t} - ({\bf b}\boldsymbol{\cdot \nabla}) {\bf u} \right] = \nabla^2 {\bf b} + \cos \phi\frac{\partial {\bf u}}{\partial x} + \sin \phi\frac{\partial {\bf u}}{\partial y}, \label{IE}\\
\mathrm{Pr}\frac{\mathcal{D} {\bf \theta}}{\mathcal{D} t} = u_z + \nabla^2 \theta,\label{EE}\\
\boldsymbol{\nabla \cdot} {\bf{u}} = 0,~~~\boldsymbol{\nabla \cdot} {\bf{b}} = 0, \label{DF}
\end{gather}
\end{subequations}
where \(\mathcal{D}/\mathcal{D} t \equiv \partial /\partial t + ({\bf u}\boldsymbol{\cdot \nabla})\) is the material derivative and \({\bf{\hat{k}}}\) is the vertical unit vector. The symbols $\bf{u}$ and $\bf{b}$ respectively represent the convective velocity and induced magnetic fields with components  $(u_x, u_y, u_z)$ and $(b_x, b_y, b_z)$. The scalar fields $p$ and $\theta$ represent the reduced kinematic pressure and the deviation of the temperature field from the steady conduction profile. The non-dimensionalization of the equations~(\ref{ME})-(\ref{DF}) are done using the units $l$, $l^2/\nu$, \(\Delta T \nu/\kappa\) and \(B_0\nu/\eta\) for length, time, temperature and magnetic field respectively.  The non-dimensionalization process leads to four dimensionless parameters, namely, the Rayleigh number (\(\mathrm{Ra} = \alpha g \beta l^4/\nu \kappa\), the buoyancy force with $g$ being the acceleration due to gravity), the Chandrasekhar number (\(\mathrm{Q} = B_0^2l^2/\rho_0 \nu \eta\), strength of the magnetic field), the thermal Prandtl number (\(\mathrm{Pr} = \nu/\kappa\), ratio of the thermal and viscous diffusion time scales), and the magnetic Prandtl number (\(\mathrm{Pm} = \nu/\eta\), ratio of the magnetic and viscous diffusion time scales) which governs the dynamics of the system. Here $\rho_0$ is the reference density. 

In this paper, our aim is to investigate the effect of external horizontal oblique magnetic field on the convection rolls in electrically conducting low Prandtl number fluids for which $\mathrm{Pm}$ is extremely small ($\sim 10^{-5}$). Further, we focus on the flow patterns near the onset of convection where the hydrodynamic Reynolds number $Re$ is assumed to be small. Thus, the magnetic Reynolds number $Rm$ is very small in the present study, and due to this, we simplify the mathematical model of the problem by considering the quasi-static approximation~\cite{roberts_book,knaepen_JFM:2004,yan:2019}. As a result, the equations (\ref{ME}) and~(\ref{IE}) simplify to 

\begin{subequations}
\begin{gather}
\frac{\mathcal{D} {\bf u}}{\mathcal{D} t} - \mathrm{Q} \left[ \cos \phi\frac{\partial {\bf b}}{\partial x} + \sin \phi\frac{\partial {\bf b}}{\partial y} \right] = -\pmb{\nabla} p + \mathrm{Ra}\theta {\bf \hat{k}} + \nabla^2 {\bf u},  \label{ME_Pm0} \\
\nabla^2 {\bf b} = -\left[\cos \phi\frac{\partial {\bf u}}{\partial x} + \sin \phi\frac{\partial {\bf u}}{\partial y}\right]. \label{IE_Pm0}
\end{gather}
\end{subequations}

The horizontal plates are considered to be stress-free, perfect conductors of heat and electricity which leads to the following conditions: 

\begin{equation}
v_3 = \frac{\partial v_1}{\partial z} = \frac{\partial v_2}{\partial z} = \theta = b_3 = \frac{\partial b_1}{\partial z} = \frac{\partial b_2}{\partial z} = 0 \quad \text{at} \quad z = 0,1. \label{BC}
\end{equation}

Additionally, all convective fields are assumed to be periodic in the horizontal directions. Mathematically, this can be expressed as:
\begin{equation}
F(x+\mathrm{L_x}, y, z) = F(x, y+\mathrm{L_y}, z) = F(x, y, z), \label{BC1}
\end{equation}
where $F=({\bf u}, {\bf b}, \theta)$ and $\mathrm{L_x}$, $\mathrm{L_y}$ are the periods in the x, y directions, respectively. Thus, the mathematical model of the physical system under consideration in the presence of the oblique horizontal magnetic field is given by the equations (\ref{EE}) - (\ref{IE_Pm0}) along with the supplementary conditions (\ref{BC}) and (\ref{BC1}). In this paper, we determine the critical Rayleigh number ($\mathrm{Ra}_c = \frac{27}{4}\pi^4$) for the onset of convection along with critical wave number $\mathrm{k}_c = \frac{\pi}{\sqrt{2}}$ in the absence of external magnetic field using the linear theory based on the normal mode analysis~\cite{chandra:book} and use it in the numerical simulation subsequently.

\section{Numerical Simulation Details}
We perform direct numerical simulations (DNS) of the mathematics model described in the previous section using the open-source pseudospectral code Tarang~\cite{mkv:code}. In the simulation, the independent fields are expanded using a set of orthogonal basis functions, either $\lbrace e^{i(lk_x x+mk_y y)} \sin(n \pi z) \mid l, m, n = 0, 1, 2, \cdots \rbrace$ or $\lbrace e^{i(lk_x x+mk_y y)} \cos(n \pi z) \mid l, m, n = 0, 1, 2, \cdots \rbrace$, compatible with the boundary conditions. The wave numbers along the $x$-axis and $y$-axis are denoted by $k_x$ and $k_y$, respectively. Thus, the vertical velocity, vertical vorticity, and temperature fields are expanded in terms of Fourier amplitudes or modes $W_{lmn}$, $Z_{lmn}$, and $T_{lmn}$ as

\[
\begin{aligned}
\left(\begin{array}{c}
u_z(x,y,z,t) \\
\omega_z(x,y,z,t) \\
\theta(x,y,z,t)
\end{array}\right) 
= \sum_{l,m,n} 
\left(\begin{array}{c}
W_{lmn}(t) \\
Z_{lmn}(t) \\
T_{lmn}(t)
\end{array}\right)
e^{i(lk_x x + mk_y y)}
\left(\begin{array}{c}
\sin(n\pi z) \\
\cos(n\pi z) \\
\sin(n\pi z)
\end{array}\right).
\end{aligned}
\]

We perform detailed direct numerical simulations for three different horizontal aspect ratios given by $\Gamma = {\mathrm{L_x}}/{\mathrm{L_y}}= 1,~\frac{1}{2}$ and $2$. The grid resolution is chosen according to the aspect ratio. For instance, grid resolutions of $32^2$ or $64^3$ were used for $\Gamma=1,2$, while for $\Gamma=\frac{1}{2}$, grid resolutions of $32 \times 64 \times 32$ or $64 \times 128 \times 64$ were used. The time advancement is done using the fourth-order Runge-Kutta (RK4) method. The Courant-Friedrichs-Lewy (CFL) condition is applied to ensure numerical stability, dictating the maximum allowable time step based on the grid spacing and flow speed. The CFL condition is expressed as:
     \[
     \delta t \leq \frac{\text{min}(\Delta x, \Delta y, \Delta z)}{\text{max velocity}},
     \]
where \( \delta t \) is the time step, and \( \Delta x, \Delta y, \Delta z \) are the grid spacings in the $x$, $y$, and $z$ directions, respectively. The chosen time step \( \delta t = 10^{-3} \) in this study satisfies the CFL condition. Additionally, a parameter $r$, known as the reduced Rayleigh number, defined as $r = \mathrm{Ra}/\mathrm{Ra_c}$, is used for convenience.
     
\section{Oblique magnetic field and Flow patterns} 
It is well known that the electrically conducting low Prandtl number fluids allow convection in the form of simple cellular patterns including steady parallel rolls (SPR) close to the onset of convection~\cite{busse:JFM_1972,croquette1:Contemp.Phys_1989,thual:1992,pal:2009,mishra:EPL_2010,pal:2002,nandu:2015,ghosh:2017,mandal_POF:2023,sharma:PRF_2022}. In this paper, we investigate the effect of oblique horizontal external magnetic field on these flow patterns. The investigation is done by varying the angle $\phi$ between the axis of the steady parallel rolls (taken as $x$-axis) and the direction of the magnetic field in the interval $[0^{\mathrm{o}}, 90^{\mathrm{o}}]$ for three horizontal aspect ratios $\Gamma = {k_y}/{k_x} = 1, \frac{1}{2}$ and $2$. The simulations are performed in a box of dimensions $\mathrm{L_x} \times \mathrm {L_y}\times 1$, where $\mathrm {L_x} = {2\pi}/{k_x}$ and $\mathrm {L_y} = {2\pi}/{k_y}$. 

\subsection{Flow dynamics in a square cell ($\Gamma = 1$)}
Here we study the flow patterns in a square simulation box given by $\mathrm{L_x} = \mathrm{L_y}$ with $k_x = k_y = k_c = \pi/\sqrt{2}$. We start the simulation for $\mathrm{Pr} = 0.1$. In the absence of the magnetic field, the convection in the form of stationary two dimensional rolls along the $x$-axis or $y$-axis is observed close to the onset depending on the choice of initial conditions. We use the initial conditions for which the 2D rolls flow pattern along the $x$-axis, called steady parallel rolls (SPR) is observed at the onset.  

\subsubsection{Inhibition of convection by magnetic field}
\begin{figure}[h]
\centering
\includegraphics[scale = .5]{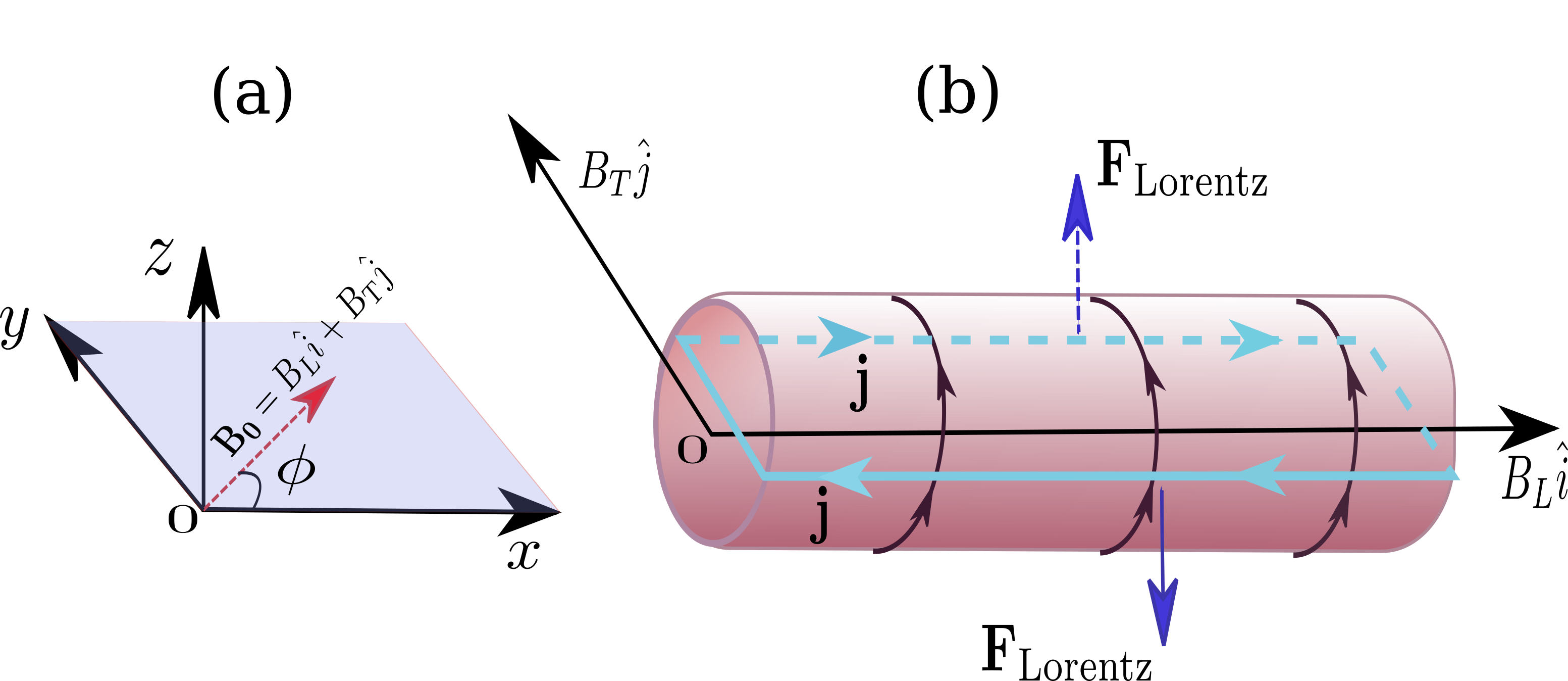}
\caption{Schematic diagrams showing: (a) Oblique horizontal magnetic field of magnitude $B_0$ making an angle $\phi$ with the $x$-axis having longitudinal and transversal components $B_L$ and $B_T$ respectively. (b) Action of Lorentz force ${\bf F}_\mathrm{Lorentz}$ on the convection rolls. The cylinder shows a typical convection roll, while the directed solid black curves on it represent the velocities of the fluid particles. The directed cyan curve shows current density ${\bf j}$ and blue arrows show the Lorentz forces acting on the fluid particles. }
\label{Schematic}
\end{figure}

\begin{figure}[h]
\centering
\includegraphics[scale = .4]{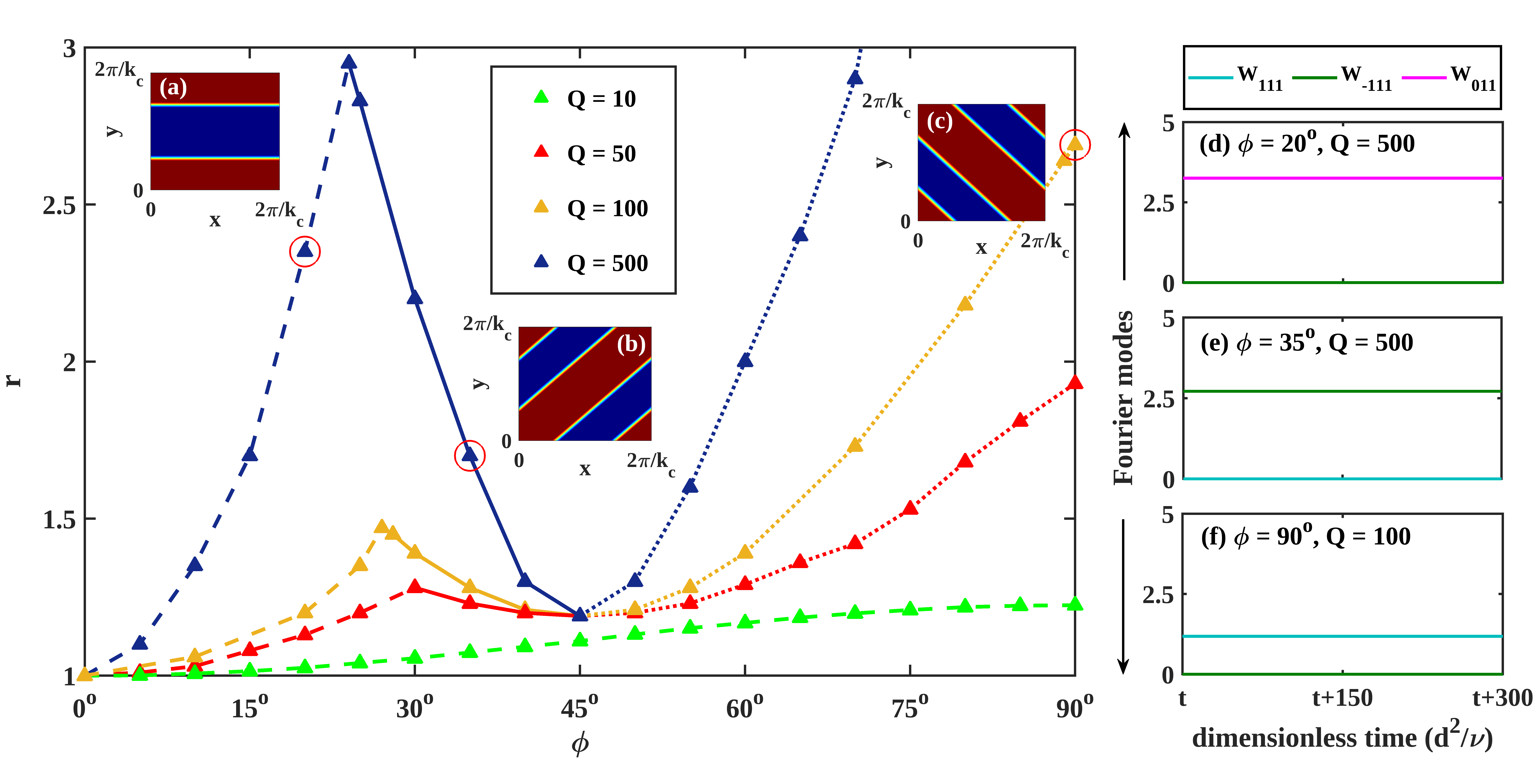}
\caption{Left panel shows the variation of $r$ for the restart of convection as a function of $\phi$ for $\mathrm{Pr} = 0.1$ and four values of $\mathrm{Q}$ as determined from the DNS. The green curve for $\mathrm{Q} = 10$ is monotonically increasing in the entire range of $\phi$. Each of the other three curves can be divided into three parts. The dashed lines represent the monotonically increasing first part. The solid and dotted parts of the curves respectively represent the monotonically decreasing and increasing parts. The insets (a), (b), and (c) respectively show the SPR, SOR$^+$ and SOR$^-$ flow patterns corresponding to the points marked by the red circles on the dashed, solid, and dotted curves. (d), (e), and (f) show the time evolution of the important modes corresponding to (a), (b), and (c) respectively.}
\label{Onset_Fig1}
\end{figure}

Slightly above the onset of convection, as the convective motion in the form of SPR sets in, the external oblique horizontal magnetic field ${\bf B}_0 = (B_0 \cos \phi, B_0 \sin \phi,0) = B_L\hat{i} + B_T\hat{j}$ is switched on and the resulting Lorentz force severely affects the convection. The effect of the magnetic field on the SPR flow patterns can be understood by looking at the effects of the components of it along $x$ and $y$ directions separately. As argued in the paper~\cite{fauve_et.al:JPL_1981}, the longitudinal component $B_L$ of the magnetic field will not affect the convection in the form of 2D rolls along the $x$-axis. On the other hand, the electric current ${\bf j} = {\bf{u}}\times B_T\hat{j}$ generated by the transverse component $B_T$ of the magnetic field will produce the Lorentz force ${\bf F}_\mathrm{Lorentz} \sim {\bf j}\times B_T\hat{j}$ inhibiting the convective motion in the form of two dimensional rolls as shown in the figure~\ref{Schematic}. Thus, for fixed choices of $\mathrm{Q}$ and $\phi ~(\neq 0)$, a larger buoyancy force (larger $r$) is necessary to restart the convection. The variation of $r$ with $\phi$ for the restart of convection has been shown in the left panel of the figure~\ref{Onset_Fig1} for four different $\mathrm{Q}$. For a lower value of $\mathrm{Q} ~(= 10)$, the curve showing the value of $r$ increases monotonically with $\phi$. This occurs due to the monotonically increasing nature of the transverse magnetic field component $B_T$ which plays the key role in suppressing the convection rolls along the $x$-direction. In this case, the convection restarts in the form of SPR. 

Interestingly, for higher $\mathrm{Q}$, the curves are not monotonic in the entire interval $[0^{\mathrm{o}}, 90^{\mathrm{o}}]$. The interval can be divided into three distinct parts, in each of which the curves respectively represented by dashed, solid, and dotted lines are monotonic. In the first part, the curve is monotonically increasing, then decreasing in the second part and again increasing in the last part. We observe that as crosses the curves through the first part, SPR flow patterns are observed as shown in the inset (a) of the figure~\ref{Onset_Fig1}. However, as $r$ crosses the curves in the second and third parts for different values of $\phi$, steady oblique rolls (SOR$^+$) flow patterns making an angle $45^\circ$ with the $x$-axis as shown in the inset (b) are observed. As a result, the angle between the oblique rolls and the external magnetic field changes to $|\phi - 45^{\mathrm{o}}|$. The value of the transverse component of the external magnetic field with respect to the oblique rolls gets reduced substantially. Accordingly, the value of $r$ for the restart of convection gets reduced and corresponding curves become monotonically decreasing. The curves reach a minima for $\phi = 45^{\mathrm{o}}$ and once again increase for $\phi > 45^{\mathrm{o}}$ with the increase of the angle between the oblique rolls and the magnetic field. We note here that when $\phi = 90^{\mathrm{o}}$, for larger values of $\mathrm{Q}$, steady oblique rolls (SOR$^-$) patterns making an angle $135^{\mathrm{o}}$ with the $x$-axis are observed, at the restart value of $r$. Corresponding flow patterns have been shown in the inset (c) of the left panel of the figure~\ref{Onset_Fig1} which is the same as the ones reported in~\cite{sharma:2023}. 

Next to identify the modes associated with the flow patterns shown in the insets (a), (b) and (c) of the figure~\ref{Onset_Fig1} we plot the time evolution of the important modes in the figures~\ref{Onset_Fig1}(d), (e) and (f).  It is clearly seen that the modes $W_{011}$, $W_{-111}$ and $W_{111}$ determine the nature of the flow patterns presented in (a), (b), and (c) of the figure~\ref{Onset_Fig1} respectively. For more details about the dominant Fourier modes at the point of restart of convection in the presence of an external oblique magnetic field, we explore the region given by $0^{\mathrm{o}} \leq \phi \leq 90^{\mathrm{o}}$ for five  $\mathrm{Q}$ with $\mathrm{Pr} =0.1$. The results are presented in the Table~\ref{Onset_Tab1}. 

\begin{table}[h]
\caption{The dominant Fourier modes for various values of $\mathrm{Q}$.}
\begin{tabularx}{1\textwidth} { 
   >{\centering\arraybackslash}X 
   >{\centering\arraybackslash}X 
   >{\centering\arraybackslash}X
   >{\centering\arraybackslash}X
   >{\centering\arraybackslash}X
   >{\centering\arraybackslash}X
   }
 \hline
 \hline
 DFM & $\phi(\mathrm{Q}=10)$ & $\phi(\mathrm{Q}=20)$ & $\phi(\mathrm{Q}=50)$ & $\phi(\mathrm{Q}=100)$ & $\phi(\mathrm{Q}=500)$\\
\hline
$\mathrm{W}_{011}$  & $0^\circ-90^\circ$ & $0^\circ-90^\circ$ & $0^\circ-30^\circ$  &	 $0^\circ-27^\circ$	& $0^\circ-24^\circ$      \\
$\mathrm{W}_{-111}$ & $-$        & $-$        & $31^\circ-89^\circ$ & $28^\circ-89^\circ$	&  $25^\circ-89^\circ$    \\ 
$\mathrm{W}_{111}$  & $-$        & $-$        & $90^\circ$      & $90^\circ$	&   $90^\circ$      \\
\hline
\end{tabularx}
\label{Onset_Tab1}
\end{table} 
The table suggests that the system exhibits one of the three primary modes, namely, $W_{011}$, $W_{-111}$, and $W_{111}$ at the onset of convection after the application of the external oblique magnetic field. For weaker magnetic field, $W_{011}$ is the dominant in the entire interval $0^{\mathrm{o}}\leq \phi \leq  90^{\mathrm{o}}$. While, with the increase of $\mathrm{Q}$, the interval of the dominance of the mode $W_{011}$ gradually reduced. On the other hand, the interval of $\phi$ for the dominance of the mode $W_{-111}$ gradually increases. Interestingly, the mode $W_{111}$ is dominant for higher $\mathrm{Q}$ values, only at $\phi = 90^{\mathrm{o}}$, which is consistent with the results of Sharma et al.~\cite{sharma:2023}. Note that when one of the aforementioned modes is dominant, the other two modes do not appear. In the following subsection, we present the bifurcation scenario of the system near the onset of convection obtained by increasing the reduced Rayleigh number.

\subsubsection{Bifurcation structure for higher $r$}
As discussed above, with the application of the oblique horizontal magnetic field close to the onset, convection in the form of SPR is suppressed. Convection restarts for a larger value of $r$ manifesting itself as stationary convection exhibiting any one of the SPR, SOR$^+$, and SOR$^-$ flow patterns depending on the choice of the parameters. Out of which we have noted that SOR$^-$ flow patterns appear at the onset only for $\phi = 90^{\mathrm{o}}$ when the stronger magnetic field is applied. The bifurcation structure associated with the SOR$^-$ flow pattern with the increase of $r$ has been discussed in detail by Sharma et al.~\cite{sharma:2023}. Here we investigate the bifurcation structure associated with the SPR and SOR$^+$ by increasing the value of $r$. 

\begin{figure}[h]
\centering
\includegraphics[scale = .4]{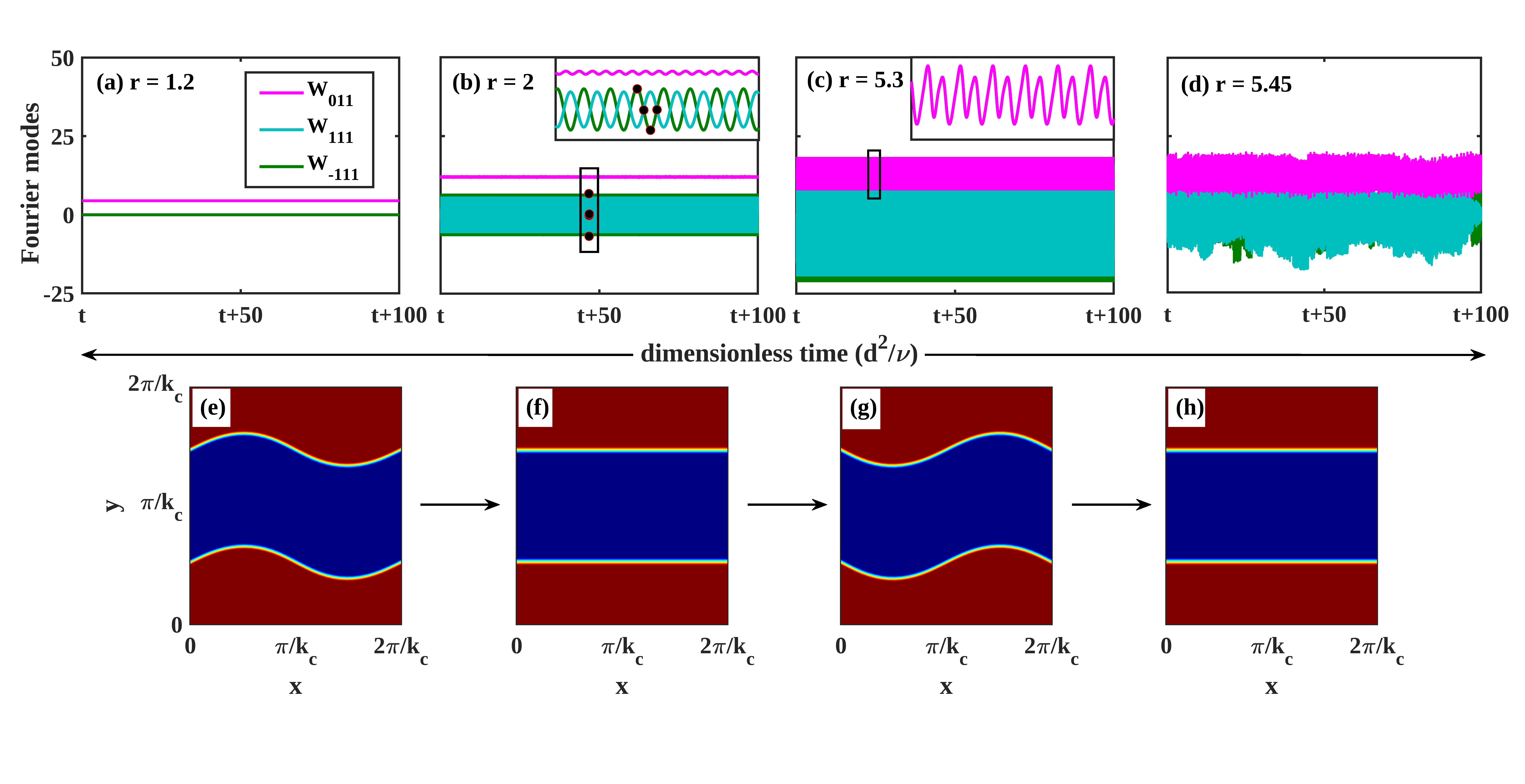}
\caption{The temporal evolution of the Fourier modes $W_{011}$, $W_{111}$ and $W_{-111}$ for (a) SPR, (b) PPR, (c) $\mathrm{P_2PR}$, and (d) CPR flow patterns for $\mathrm{Q} = 30$, $\mathrm{Pr} = 0.1$, $\phi = 30^\circ$, and $\Gamma=1$. The insets in (b) and (c) provide an enlarged view of the marked regions. Panels (e)-(h) illustrate the pattern dynamics at four different time instants marked by black dots in (b) for the PPR flow pattern.}
\label{AR1_Fig1}
\end{figure}

For that purpose first, we consider $\mathrm{Q} = 30$ and $\phi = 30^\circ$, where the SPR pattern is observed when convection restarts at $r = 1.17$. The value of $r$ is increased further till the chaotic solutions are obtained. The time evolution of the three significant modes $W_{011}$, $W_{111}$ and $W_{-111}$ are shown in the figures~\ref{AR1_Fig1}(a)-(d) for four values of $r$. For $r =1.2$, SPR solution is observed for with $W_{011}$ as the dominant mode, while the modes $W_{111}$ and $W_{-111}$ remain zero as shown in the figure~\ref{AR1_Fig1}(a). The pattern is stationary and similar to the one shown in the inset (a) of the figure~\ref{Onset_Fig1}. As $r$ is increased to $2$, we detect the emergence of a periodic solution for which the modes $W_{011}, W_{111}, W_{-111}$ oscillate over time, with $W_{011}$ remaining the dominant one (figure~\ref{AR1_Fig1}(b)). The resulting flow patterns are dynamic in nature with varied wavyness and oscillate periodically while remaining oriented along the $x$-axis. The corresponding pattern dynamics are shown in the figures~\ref{AR1_Fig1}(e) - (h) and are called periodic parallel rolls (PPR).  Further increment in r reveals the manifestation of period-2 parallel rolls $(\mathrm{P_2PR})$  followed by chaotic parallel rolls (CPR). The time evolution of the modes $W_{011}$, $W_{111}$ and $W_{-111}$ corresponding to these solutions are shown in the figures~\ref{AR1_Fig1}(c) and (d). The related pattern dynamics closely resemble to those of PPR. Thus, the route to chaos is period doubling in this case which is similar to the experimental observation reported in~\cite{fauve_prl:1984} for weaker magnetic field.

\begin{figure}[h]
\centering
\includegraphics[scale = .4]{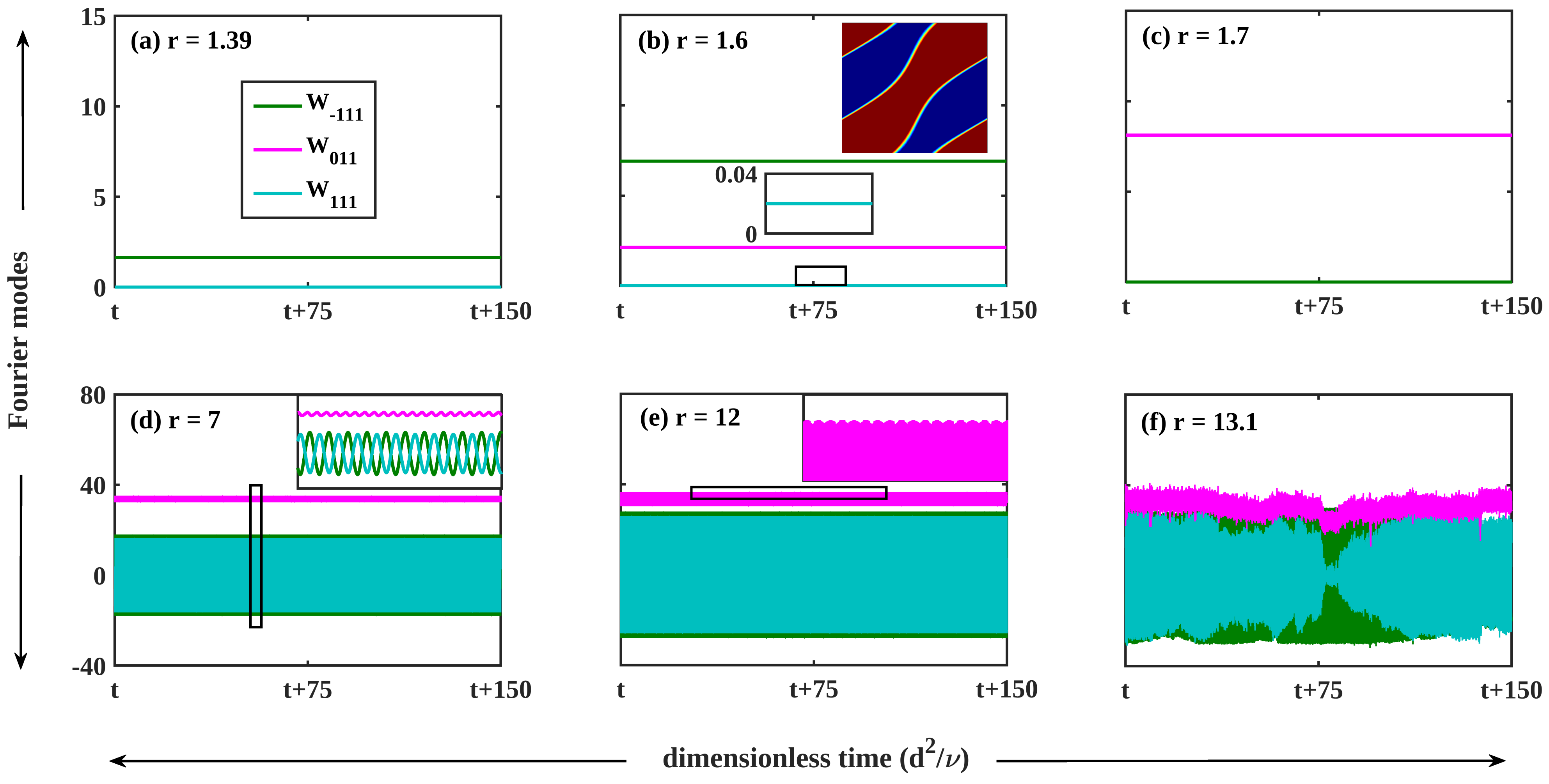}
\caption{The time evolution of the Fourier modes for (a) SOR$^+$, (b) SOWR$^+$, (c) SPR, (d) PPR, (e) QPPR, and (f) CPR flow patterns for the parameters $\mathrm{Q} = 100$, $\mathrm{Pr} = 0.1$, and $\phi = 30^\circ$ for $\Gamma=1$. The insets in (b), (d), and (e) provide enlarged views of the marked regions.}
\label{AR1_Fig2}
\end{figure}

\begin{figure}[h]
\centering
\includegraphics[scale = .35]{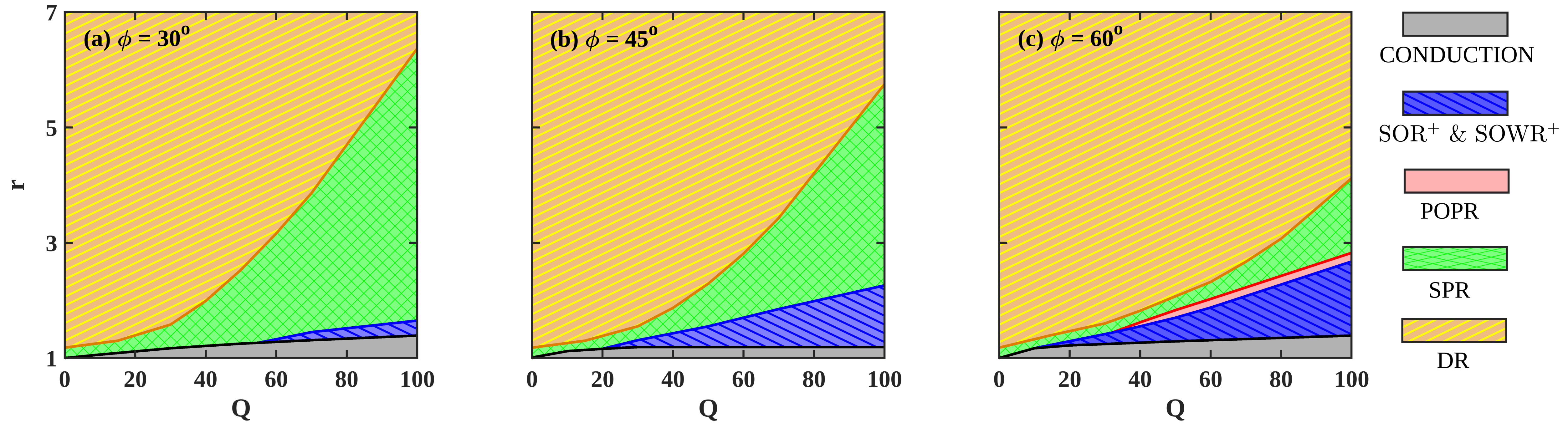}
\caption{Two-parameter diagrams showing different flow regimes on the $\mathrm{Q} - r$ plane for three values of $\phi$ with $\mathrm{Pr} = 0.1$ and $\Gamma = 1$.}
\label{AR1_Fig3}
\end{figure}

\begin{figure}[h]
\centering
\includegraphics[scale = .35]{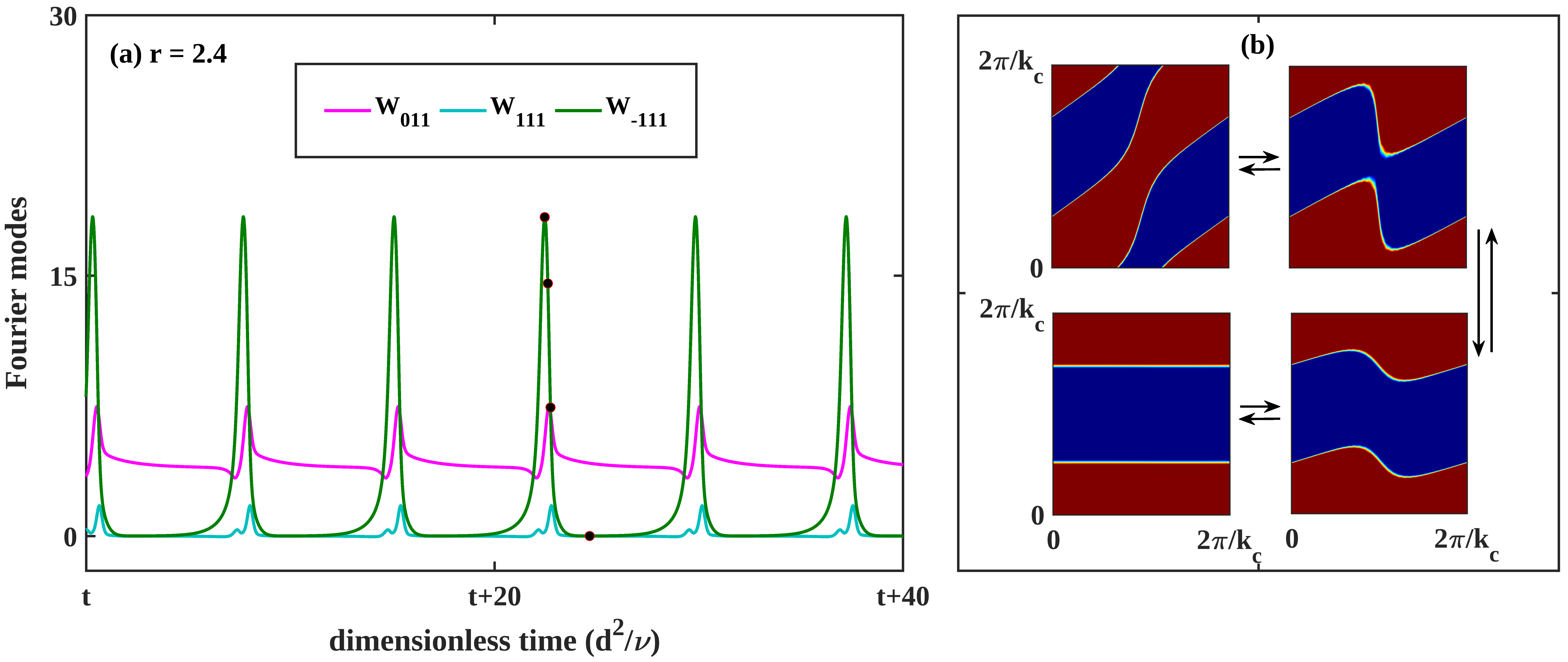}
\caption{(a) The time evolution of Fourier modes  $W_{011}, W_{111}$ and  $W_{-111}$ corresponding to the Periodic Oblique-Parallel Roll (POPR) solution obtained for $\mathrm{Pr} = 0.1$, $\phi=60^\circ$, $\mathrm{Q} = 80$, $r = 2.4$ and $\Gamma=1$. (b) Isotherms for $z = 0.5$ showing the flow patterns corresponding to the instants marked by black dots in (a).}
\label{AR1_Fig4}
\end{figure}

Now we apply a stronger magnetic field by choosing $\mathrm{Q} = 100$ for $\phi = 30^{\mathrm{o}}$, we observe convection restarts at $r = 1.39$ in the form of SOR$^+$. The time evolution of the important modes corresponding to the SOR$^+$ flow pattern is shown in figure~\ref{AR1_Fig2}(a). Note that for SOR$^+$ solution, $W_{-111} \neq 0$ and $W_{111} = W_{011} = 0$. As $r$ is increased further, we observe all modes ($W_{011}, W_{-111}, W_{111}$) becoming active and independent of time. The presence of the $W_{011}$ mode causes a slight tilt of the oblique roll towards the $x$-axis, while the $W_{111}$ mode introduces waviness into it, leading to what we term as steady oblique wavy roll (SOWR$^+$) flow pattern. The pattern dynamics associated with SOWR$^+$ is shown in the inset of the figure~\ref{AR1_Fig2}(b). The dominance of the mode $W_{011}$ is evident from the time evolution of the modes $W_{011}, W_{111}$ and  $W_{-111}$ for subsequent higher values of $r$ presented in the figures~\ref{AR1_Fig2}(c)-(f). 

The flow patterns corresponding to the stationary solution presented in the figure~\ref{AR1_Fig2}(c) is of SPR type which becomes PPR (FIG.~\ref{AR1_Fig2}(d)) followed by quasi-periodic parallel rolls (QPPR, FIG.\ref{AR1_Fig2}(e)) on increasing $r$. Subsequently, the flow becomes chaotic exhibiting chaotic parallel rolls (CPR) flow patterns. Thus, a quasi-periodic route to chaos is observed for the stronger magnetic field which is similar to the experimental observation reported in~\cite{fauve:1984}.

Now to develop a more detailed understanding of the flow dynamics in the considered parameter space we prepare two-parameter diagrams for three different $\phi$ showing qualitatively different flow regimes on the $\mathrm{Q}-r$ plane. The diagrams are shown in the figures~\ref{AR1_Fig3}(a)-(c). The figures~\ref{AR1_Fig3}(a) and (b) respectively show the flow regimes for $\phi = 30^{\mathrm{o}}$ and  $\phi = 45^{\mathrm{o}}$. In both cases, four distinct regimes are indicated by the gray, blue, green, and yellow colors respectively represent the conduction, SOR$^+$, SOWR$^+$, SPR, and dynamic rolls (DR) flow regimes consisting of different time-dependent solutions. It is observed from these two diagrams that as $\phi$ changes from $30^{\mathrm{o}}$ to $45^{\mathrm{o}}$, conduction and SPR regimes decrease, while, the SOR$^+$ and SOWR$^+$ along with DR regimes increase. Interestingly, for $\phi = 60^{\mathrm{o}}$, along with the aforementioned flow regimes, an additional flow regime named periodic oblique-parallel rolls (POPR) denoted by light red color appears between green and blue regimes. The time evolution of the important modes and pattern dynamics in this regime are shown in figure~\ref{AR1_Fig4}. This flow regime is different from other regimes mentioned above and is characterized by periodic oscillation of the parallel and oblique rolls. 

\subsubsection{Onset of oscillatory instability}

\begin{figure}[h]
\centering
\includegraphics[scale = .4]{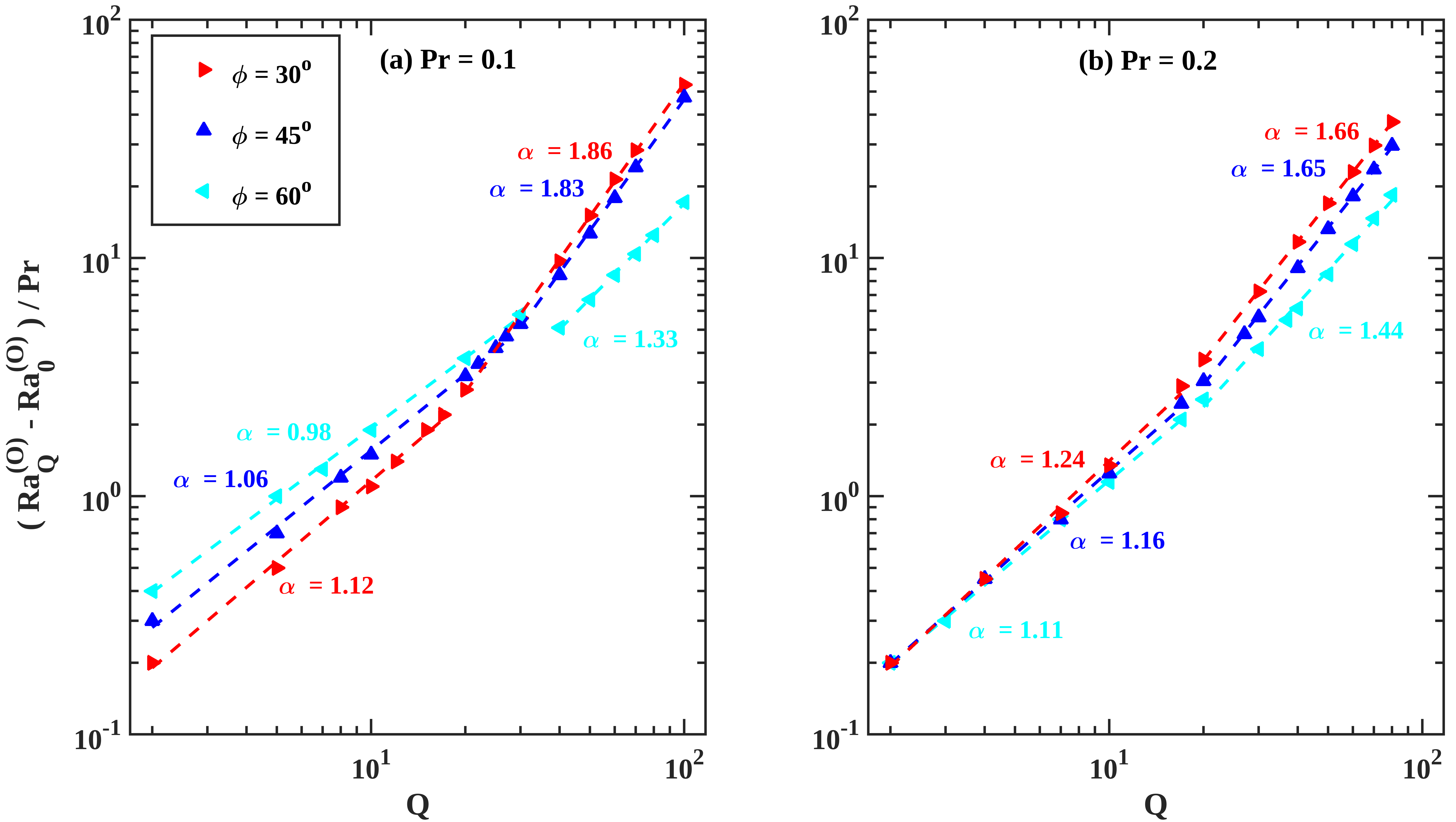}
\caption{The scaling of the onset of oscillatory instability (OI) is shown as a function of Q for Pr = 0.1 (a) and Pr = 0.2 (b) with $\Gamma=1$, considering three different magnetic field angles of $\phi=30^\circ$, $45^\circ$, and $60^\circ$. The dashed line indicates the best linear fit for the data.}
\label{Onset_OI}
\end{figure}

In the figure~\ref{AR1_Fig3}, we observe that for fixed $\phi$ and $\mathrm{Q}$,  the system transits from stationary convection regime to time-periodic regime via oscillatory instability (OI) as $r$ crosses a critical value. Here we explore the oscillatory instability of the stationary solutions in detail by considering $\phi = 30^\circ$, $45^\circ$, and $60^\circ$.  First we focus on the onset of the OI and numerically compute the quantity $(\mathrm{Ra_Q^{(O)}}-\mathrm{Ra_0^{(O)}})/\mathrm{Pr}$ as a function of $\mathrm{Q}$ at the onset of the OI for $\mathrm{Pr} = 0.1$ and $0.2$, where $\mathrm{Ra_Q^{(O)}}$ and $\mathrm{Ra_0^{(O)}}$ respectively denote the Rayleigh number for the onset of OI in the presence and absence of the external magnetic field. The variation of $(\mathrm{Ra_Q^{(O)}}-\mathrm{Ra_0^{(O)}})/\mathrm{Pr}$ with $\mathrm{Q}$ is depicted in the FIG. \ref{Onset_OI} for both $\mathrm{Pr}$ value showing scaling laws of the form $\sim \mathrm{Q}^{\alpha}$ in the considered range of $\mathrm{Q}$. We note that two different scaling are prevalent, one is for lower $\mathrm{Q}$ and the other one is for higher $\mathrm{Q}.$  On the other hand, similar scaling laws with single exponent have been reported in the presence of longitudinal magnetic field ($\phi = 0^\circ$) both experimentally~\cite{fauve:1984} and numerically~\cite{busse:1983, nandu:2015}, and only numerically in the presence of transverse magnetic field ($\phi = 90^\circ$)~\cite{sharma:2023}. A closer look at the figure reveals that both the exponents consistently decrease with $\phi$.

\begin{figure}[h]
\centering
\includegraphics[scale = .4]{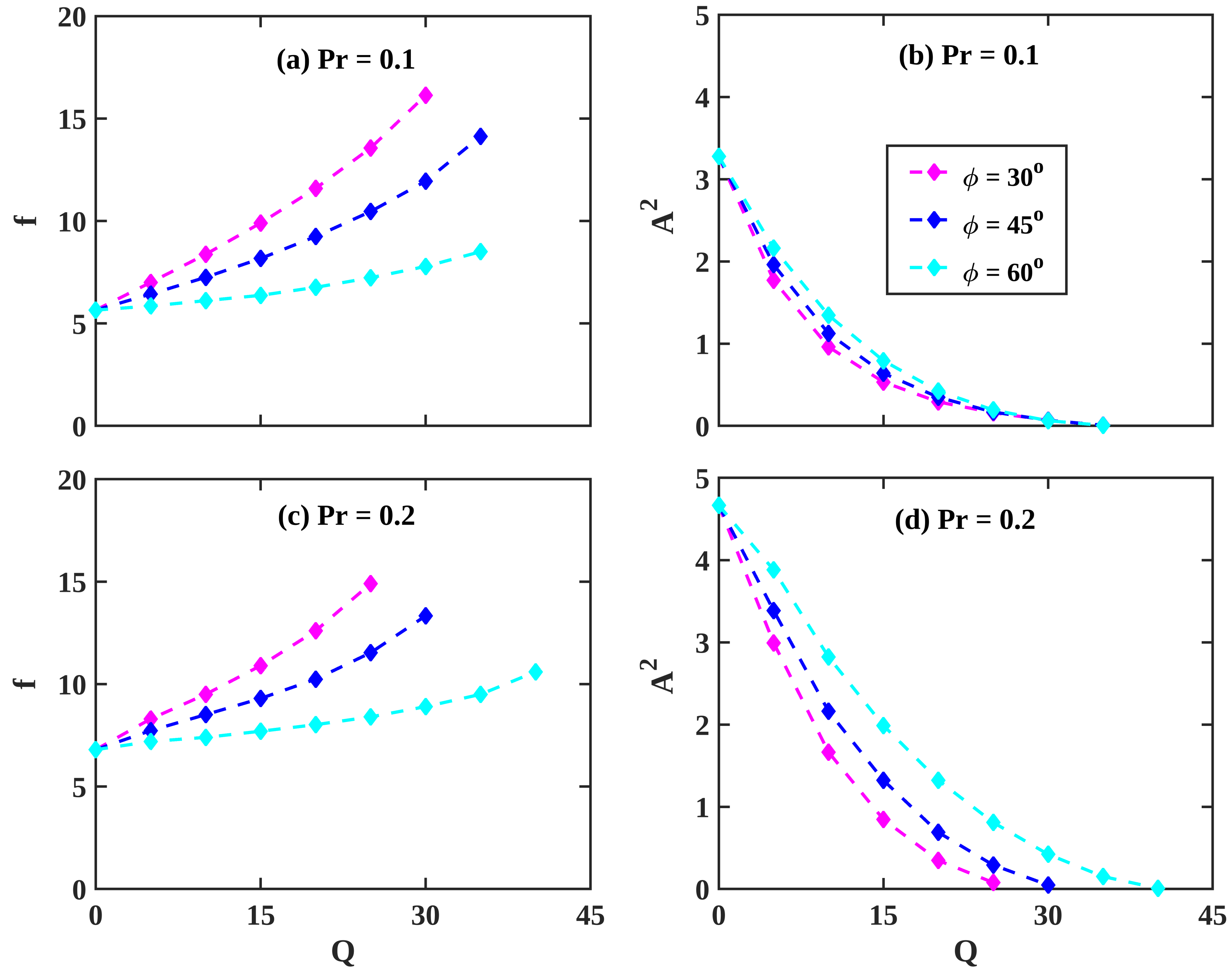}
\caption{The frequency ($f$) and amplitude ($A$) of oscillation of the periodic solution as a function of $\mathrm{Q}$  for $\Gamma=1$, three values of $\phi$ and two values of $\mathrm{Pr}$. The top ((a), (b)) and the bottom ((c), (d)) rows  are for $r = 1.75$ and $r = 2.35$ respectively.}
\label{Freq_Ampli}
\end{figure}

Next, we analyze the influence of $\mathrm{Q}$ and $\mathrm{Pr}$ on the amplitude and frequency of periodic solutions arising out of OI in detail. For this investigation, we fixed the Rayleigh number at its maximum value, where time periodic oscillatory solution (PPR)  is observed for $\mathrm{Q} = 0$, and gradually increased $\mathrm{Q}$ until the oscillations were suppressed in the system for the same reduced Rayleigh number. For $\mathrm{Pr} = 0.1$, PPR persists up to $r = 1.75$, while for $\mathrm{Pr} = 0.2$, it persists up to $r = 2.35$ in the absence of the magnetic field ($\mathrm{Q} = 0$). Therefore, we chose $r = 1.75$ and $2.35$ for $\mathrm{Pr} = 0.1$ and $0.2$, respectively, and slowly increased $\mathrm{Q}$. The variation of the oscillation amplitude and frequency of the resulting periodic solutions as a function of $\mathrm{Q}$ for three values of $\phi$ are presented in the FIG. \ref{Freq_Ampli}. It is observed from the figures~\ref{Freq_Ampli}(a) and (c) that the frequency of oscillation ($f$) gradually increases with $\mathrm{Q}$ for the three considered values of $\phi$. However, the rate of increase of $f$ with $\mathrm{Q}$ decreases as the value of $\phi$ increases. On the contrary, the amplitude of oscillations ($A$) gradually dies down with $\mathrm{Q}$ in all the cases. This observation is consistent with the experimental observation reported in Fauve et al.~\cite{fauve_et.al:JPL_1981} in the presence of longitudinal as well as transverse magnetic fields. It has been experimentally shown in~\cite{fauve_et.al:JPL_1981} that both longitudinal and transverse horizontal magnetic fields suppress the amplitude of oscillation. As a result, the oblique horizontal magnetic field also suppresses the amplitude of oscillation in the present case as shown in the figure~\ref{Freq_Ampli}(b) and (d). On the contrary, the frequency of oscillation increases at a high rate with the strength of the longitudinal magnetic field but decreases slowly in the presence of the transverse horizontal magnetic field. In the present case, a combination of longitudinal and transverse horizontal magnetic fields is applied.  The strength of the longitudinal magnetic field decreases, while that of the transverse magnetic field increases with $\phi$. Thus, as evident from the figures~\ref{Freq_Ampli}(a) and (c), the rate of the growth of frequency substantially decreases with the increase of $\phi$ due to the growing influence of the transverse magnetic field.  Now, we move on to investigate the effect of $\phi$ on the heat transfer properties of the fluid. 

\subsubsection{Effect on the heat transfer property}
Here we investigate the effect of an oblique horizontal magnetic field  on the heat transfer property of the fluids measured in terms of the global quantity called Nusselt number ($Nu$) given by
\begin{equation}
Nu = 1 + \mathrm{Pr}^2\langle v_3\theta\rangle. 
\end{equation}
Note that $Nu$ measures the ratio of average convective to conductive heat transfer in the system and the symbol $\langle \cdot\rangle$ denotes the time as well as space average over the computational domain.

\begin{figure}[h]
\centering
\includegraphics[scale = .4]{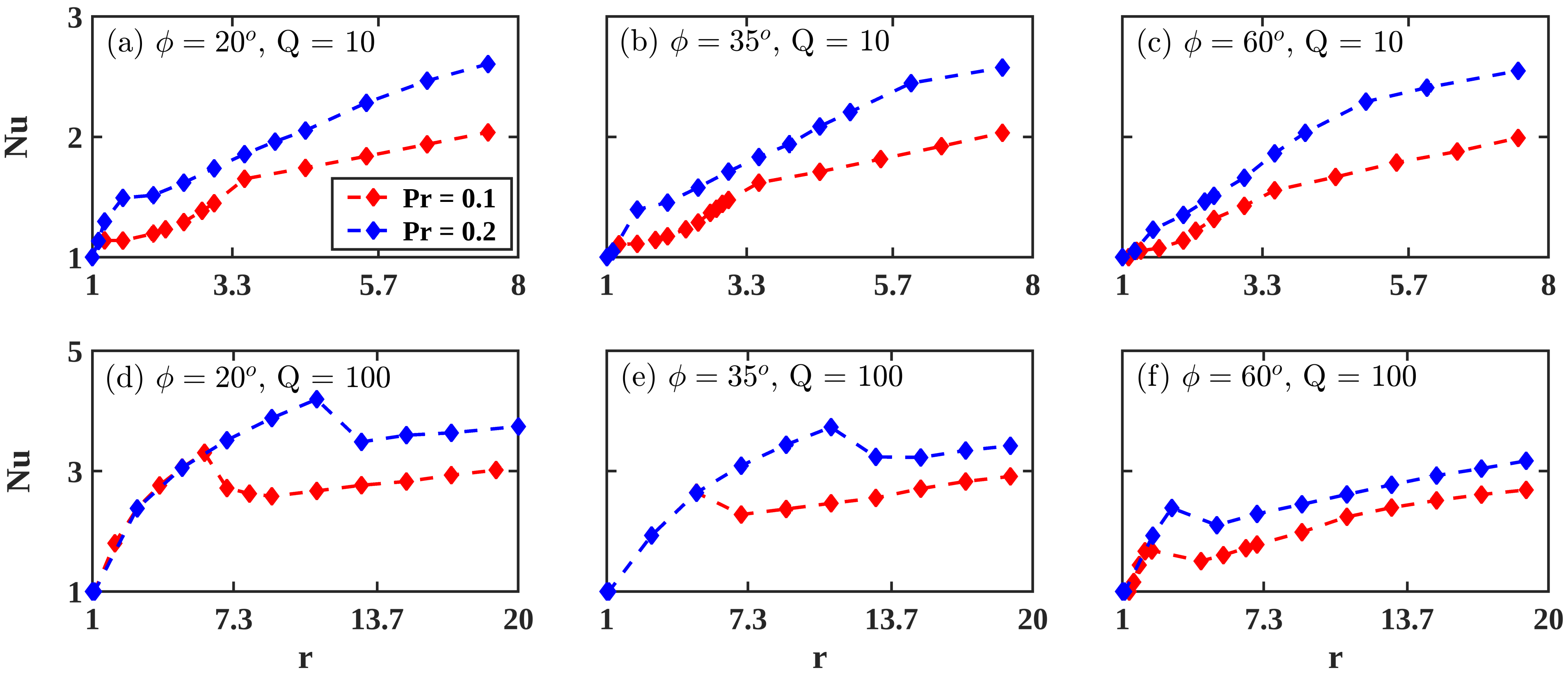}
\caption{(a)-(f) The variation of the Nusselt number ($\mathrm{Nu}$) as a function of $r$ for  $\Gamma=1$, two values of $\mathrm{Pr}$, two values of $\mathrm{Q}$, and three values of $\phi$.}
\label{Q_Nu}
\end{figure}

Figure~\ref{Q_Nu} shows the effect of the horizontal oblique magnetic field on the heat transfer property for two different Prandtl numbers. 
Top row of the FIG.~\ref{Q_Nu} is for $\mathrm{Q} = 10$, while the bottom row is for $\mathrm{Q} = 100$. From the figure~\ref{Q_Nu}, it is clearly observed that for both the Prandtl numbers, the heat transfer is inhibited with the increase of $\phi$.  

\subsubsection{Flow dynamics for strong magnetic field}
We now explore the effect of a strong magnetic field on the flow. With the increase of the strength of the external magnetic field, it is observed that the stationary flow regime exhibiting SOR$^+$ flow patterns characterized by the $W_{-111}$ mode, is greatly enhanced. In addition, interestingly, we observe a new stationary flow pattern dominated by the $W_{-211}$ mode.  
\begin{figure}[h]
\centering
\includegraphics[scale = .42]{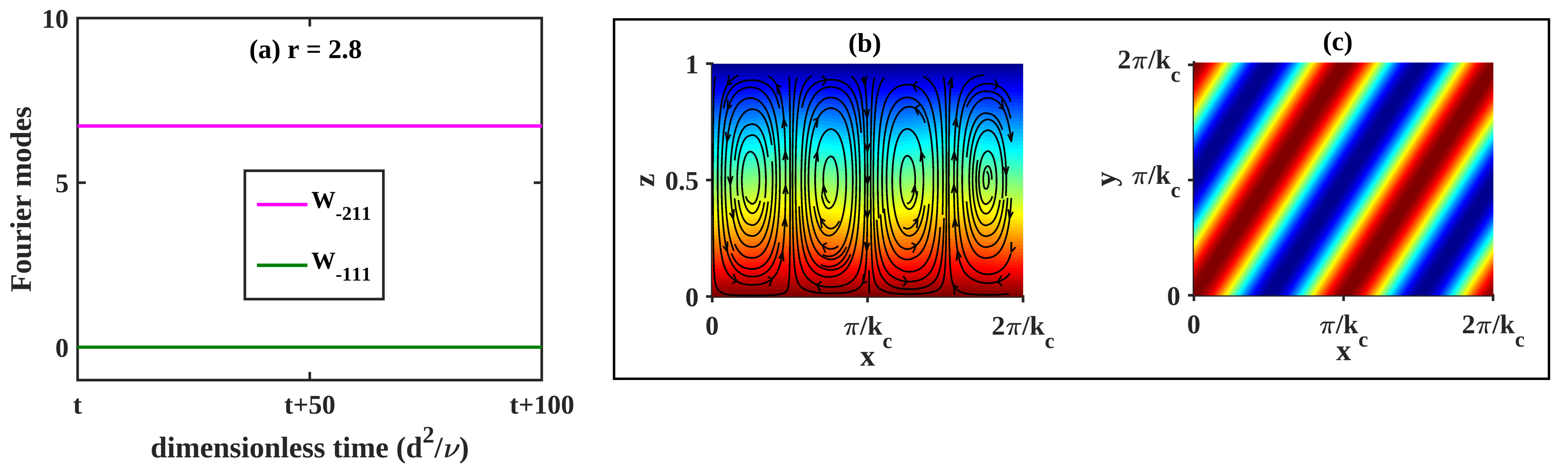}
\caption{(a) Time evolution of the modes $W_{-111}$ and $W_{-211}$, (b) streamlines on the $x-z$ plane, and (c) isotherms at the midplane $z = 0.5$ for $\Gamma=1$, $\mathrm{Pr}=0.1$, $\phi=60^{\mathrm{o}}$, $\mathrm{Q}=1000$ and $r=2.8.$}
\label{Higher_MF}
\end{figure}
The time evolution of the dominant mode, streamlines, and corresponding oblique flow patterns are shown in figure~\ref{Higher_MF}. In the following subsection, we consider a rectangular computational box and study the flow dynamics. 

\subsection{Flow dynamics in rectangular cells}
In this subsection, we present the results of our numerical investigation on the flow dynamics in two different rectangular cells given by $\Gamma = \frac{1}{2}$ and $2$. In both the cases, we perform detailed analysis for $\phi = 30^{\mathrm{o}}$, $45^{\mathrm{o}}$ and $60^{\mathrm{o}}$ with $\mathrm{Pr} = 0.1.$ 

\subsubsection{Rectangular cell with $\Gamma = \frac{1}{2}$}
Here we choose $k_x = k_c$ and $k_y = k_c/2$ to set $\Gamma = \frac{1}{2}$. Thus, the simulation box length along the $x$-axis is kept fixed to $\mathrm{L_x} = \frac{2\pi}{k_c}$ and doubled along the $y$-axis, i.e., $\mathrm{L_y} = \frac{4\pi}{k_c}$. As a result, an additional set of modes given by the odd multiples of  $k_y$ are accommodated in the expansion of different independent fields of the system. This leads to many interesting flow dynamics in the system and we discuss those subsequently.  

\begin{figure}[h]
\centering
\includegraphics[scale = .45]{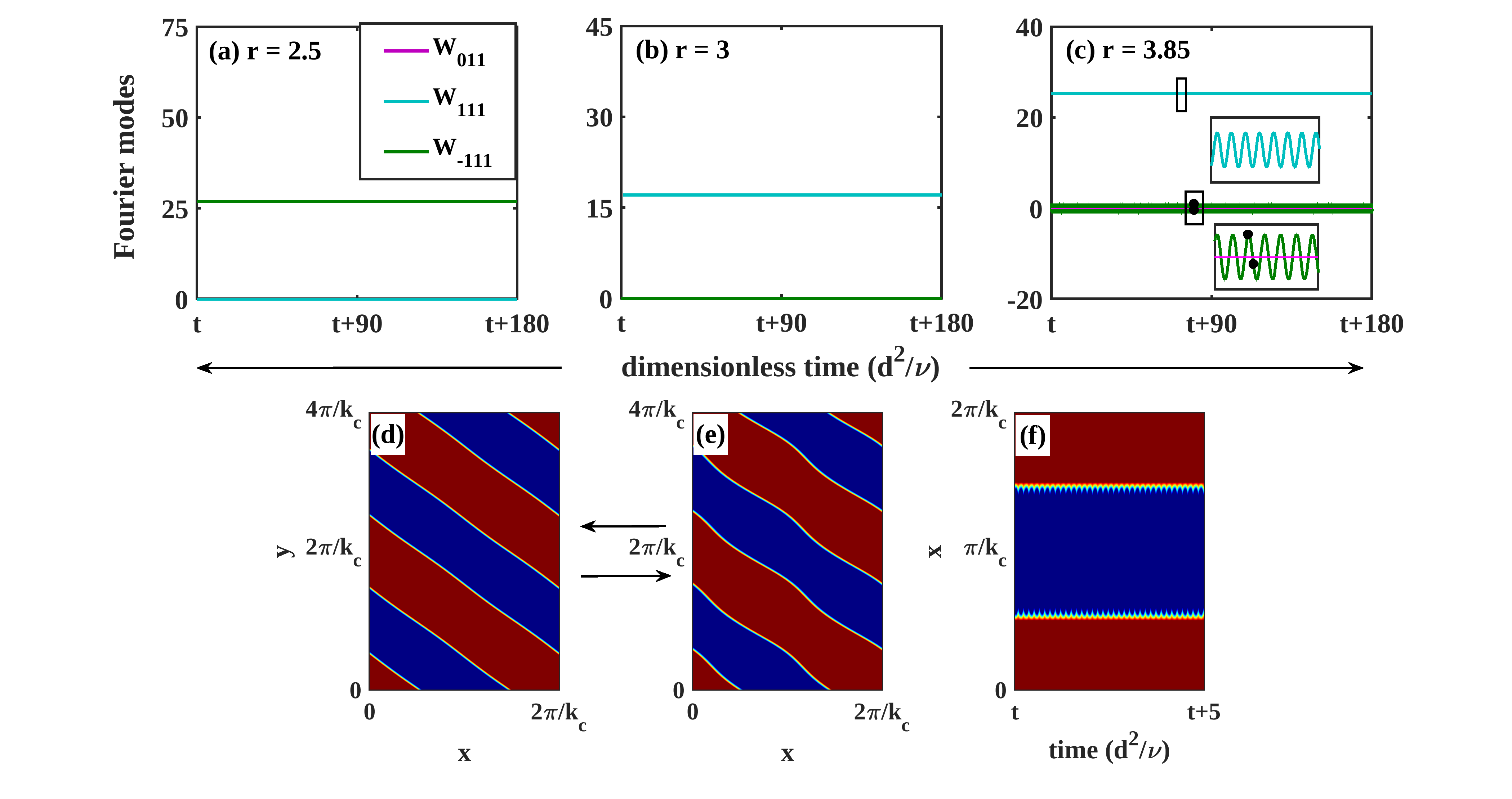}
\caption{The time evolution of the Fourier modes $W_{011}$, $W_{111}$, and $W_{-111}$  for (a) SOR$^+$, (b) SOR$^-$, and (c) POR$^-$ solutions obtained for different values of $r$ with $\mathrm{Q} = 50$, $\mathrm{Pr} = 0.1$, $\phi = 30^{\mathrm{o}}$, and $\Gamma = \frac{1}{2}$. In (c), insets shows a close-up view of the marked region. (d)-(e) Isotherms at $z =0.5$ for two different instants shown with black dots in (c). (f) Hovm\"{o}ller plot of $\theta$ corresponding to the POR$^-$ solution shown in (c) with $y = 2\pi/k_c$ and $z = 0.5$.}
\label{AR_0p5_1}
\end{figure}

\begin{figure}[h]
\centering
\includegraphics[scale = .45]{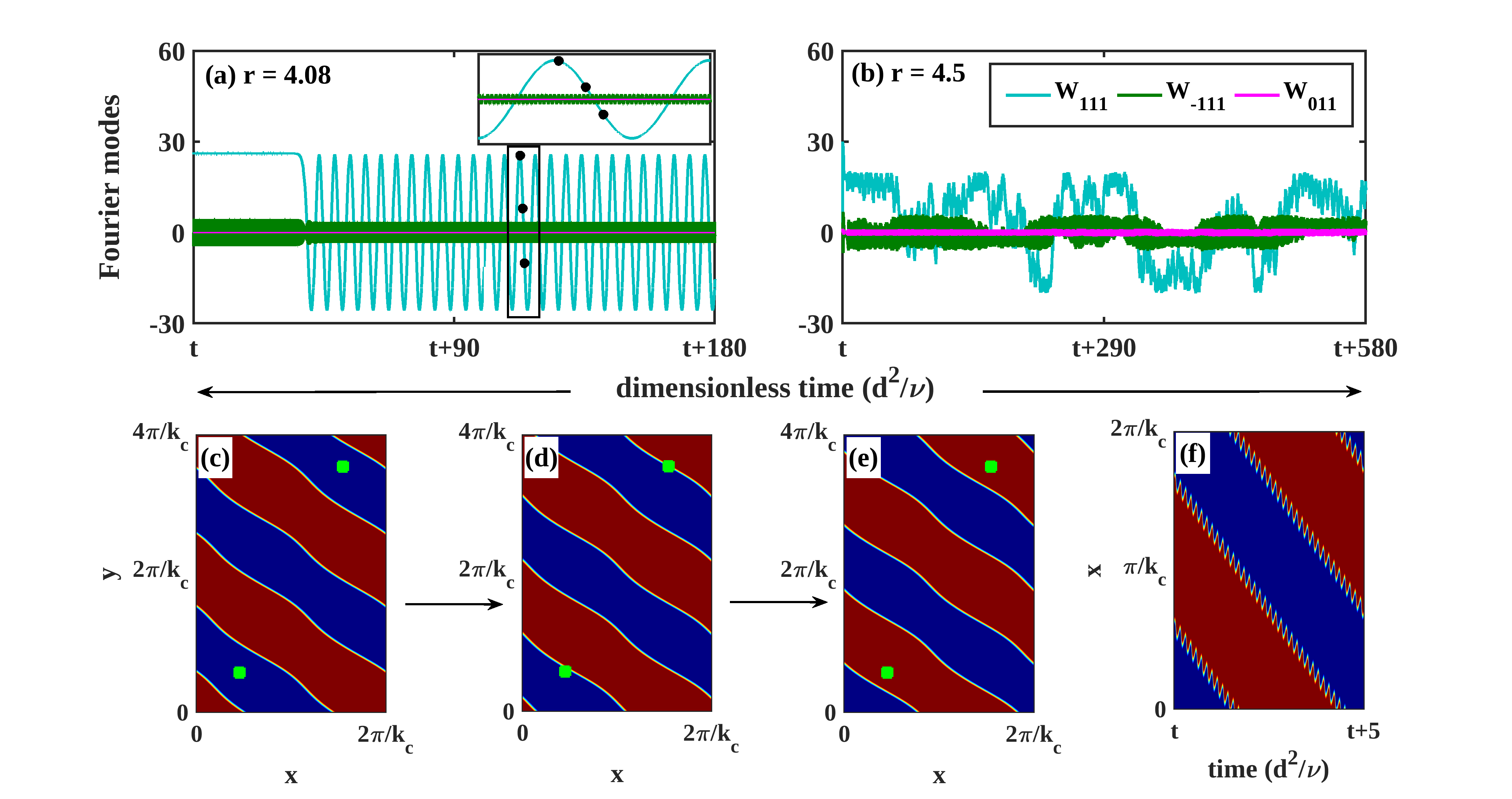}
\caption{The time evolution of the Fourier modes $W_{011}$, $W_{111}$, and $W_{-111}$  for (a) TOR$^-$ and (b) CTOR$^-$ solutions obtained for different values of $r$ with $\mathrm{Q} = 50$, $\mathrm{Pr} = 0.1$, $\phi = 30^{\mathrm{o}}$, and $\Gamma = \frac{1}{2}$. In (a), insets shows a close-up view of the marked region. (c)-(e) Isotherms at $z =0.5$ for three different instants shown with black dots in (a). (f) Hovm\"{o}ller plot of $\theta$ corresponding to the TOR$^-$ solution shown in (a) with $y = 2\pi/k_c$ and $z = 0.5$. The green squares in (c)-(e) indicate two fixed points in the domain.} 
\label{AR_0p5_2}
\end{figure}

\begin{figure}[h]
\centering
\includegraphics[scale = .35]{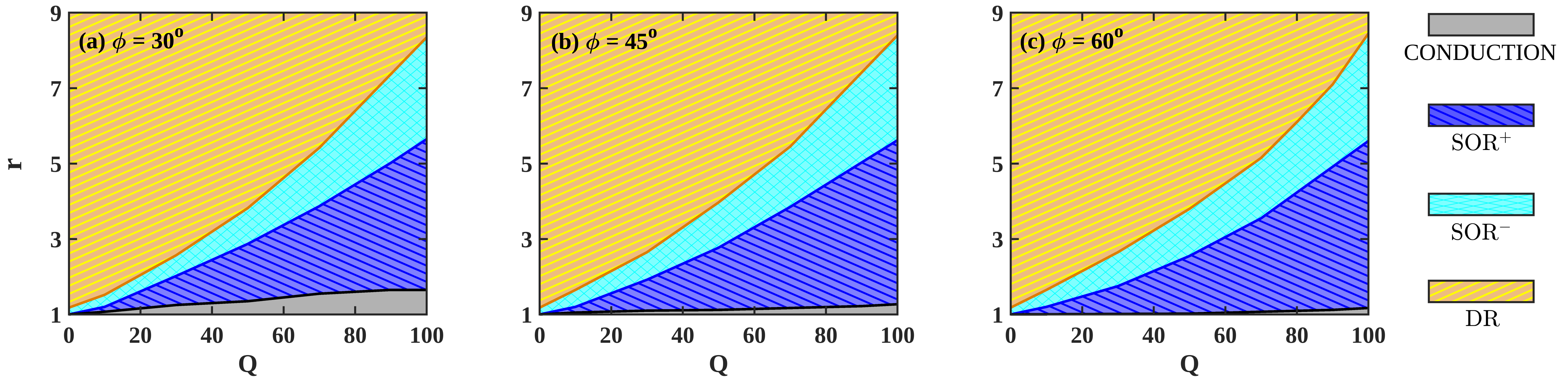}
\caption{(a) - (c) Two-parameter diagrams showing different flow regimes on the $\mathrm{Q} - r$ plane for $\mathrm{Pr} = 0.1$, $\Gamma = \frac{1}{2}$ and three values of $\phi$.}
\label{AR_0p5_3}
\end{figure}

First, we consider $\phi = 30^\circ$ and $\mathrm{Q} = 50$, and observe convection appears as SOR$^+$ at the onset for which $W_{-111}$ is the dominant mode.  The time evolution of the important Fourier modes corresponding to SOR$^+$ is shown in the figure~\ref{AR_0p5_1}(a). As $r$ increases, the system transits to the stationary oblique rolls (SOR$^-$) flow regime, dominated by the mode $W_{111}$ (Fig. \ref{AR_0p5_1}(b)). Further, an increase in $r$ leads the system to a time-dependent periodic oblique rolls (POR$^-$) flow regime. The time evolution of the important Fourier modes for POR$^-$ is shown in the figures~\ref{AR_0p5_1} (c). The isotherms computed at the midplane $z =0.5$ corresponding to the black dots marked in the figure~\ref{AR_0p5_1} (c) are shown in the figures~\ref{AR_0p5_1}(d)-(e). These time-dependent flow patterns are of standing wave type as evident from the horizontal structure of the space-time plot called Hovm\"{o}ller plot~\cite{hovmoller:1949, mandal:POF_2022} shown in the figure~\ref{AR_0p5_1}(f).  We also observe a traveling oblique rolls flow regime (TOR$^-$) as $r$ is increased further, for which the time evolution of the important Fourier modes are shown in figure~\ref{AR_0p5_2}(a). The isotherms as well as the corresponding Hovm\"oller plot for this traveling wave solution are shown in the figure~\ref{AR_0p5_2}. To understand the traveling nature of the flow patterns, we have focused on two specific points marked by green squares on the isotherms shown in the figures~\ref{AR_0p5_2}(c)-(e). The movement of the patterns with time across those marked points indicates the traveling nature of it. It is further confirmed from the diagonal orientation in the  Hovm\"{o}ller plot shown in the figure~\ref{AR_0p5_2}(f). The system eventually becomes chaotic for a higher value of $r$ exhibiting chaotic traveling oblique rolls (CTOR$^-$) flow patterns. The corresponding time evolution of the important modes is shown in the figure~\ref{AR_0p5_2}(b).

To examine the detailed effects of $\mathrm{Q}$ and $\phi$ on the flow patterns, we prepare three two-parameter diagrams showing qualitatively different flow regimes on the $r - \mathrm{Q}$ plane. 
The two-parameter diagrams shown in the figures~\ref{AR_0p5_3}(a)-(c) depict the conduction regime with gray color, the stationary SOR$^+$ and SOR$^-$ convective flow regimes with blue and cyan colors respectively, and the time-dependent dynamic regime (DR) consisting of standing as well as traveling flow patterns is represented by yellow color. From the figures it is seen that with the increment of $\phi$, the SOR$^+$ regime increases, while, the other convection regimes do not change much.

\subsection{Rectangular cell with $\Gamma = 2$}

We now consider a rectangular computational box with $\Gamma = 2$ by setting $k_x = k_c$ and $k_y = 2k_c$. In this case, the length of the simulation box along $x$-axis remains fixed to $\mathrm{L_x}= \frac{2\pi}{k_c}$, while that along the $y$-axis is halved, i.e. $\mathrm{L_y}= \frac{\pi}{k_c}$. The reduction of $\mathrm{L_y}$ by half, rules out the possibility of excitation of the waves of wavenumber $k_c$ and its odd multiples along the $y$-direction. As a result, interesting new convective patterns are observed in the considered parameter space. Here also we consider three values of $\phi$, namely, $30^\circ$, $45^\circ$, and $60^\circ$. 

\begin{figure}[h]
\centering
\includegraphics[scale = .4]{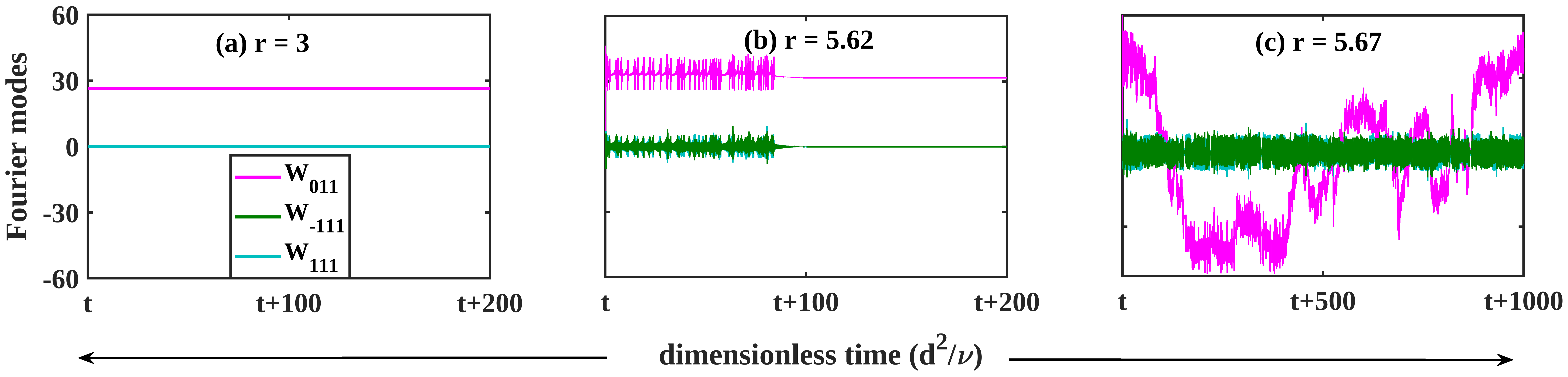}
\caption{Time evolution of the Fourier modes $W_{011}$, $W_{111}$, and $W_{-111}$ for various flow states occurring with the increment of $r$ for the fixed parameters $\mathrm{Q} = 10$, $\mathrm{Pr} = 0.1$, $\phi = 30^\circ$ and $\Gamma = 2$.}
\label{AR2_Fig1}
\end{figure}

\begin{figure}[h]
\centering
\includegraphics[scale = .4]{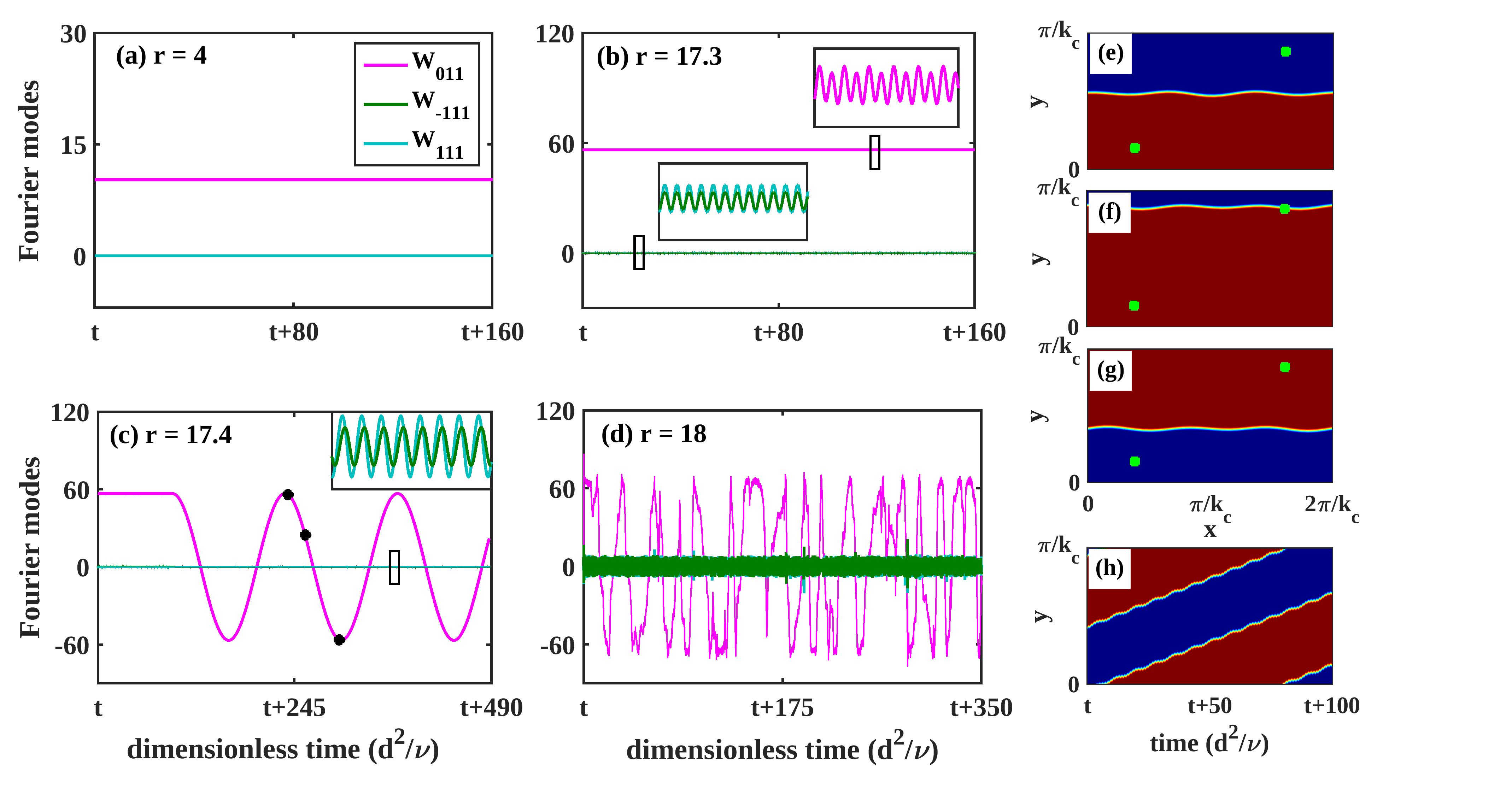}
\caption{(a) - (d) Time evolution of the Fourier modes $W_{011}$, $W_{111}$, and $W_{-111}$ for the (a) SPR, (b) PPR, (c) TPR and (d) CTPR solutions as obtained from the DNS with the increment of $r$ for fixed  $\mathrm{Q} = 30$, $\mathrm{Pr} = 0.1$, $\phi = 30^\circ$ and $\Gamma=2$. The insets in (b), and (c) provide an enlarged view of the marked regions. (e)-(g) Isotherms at $z =0.5$ corresponding to the instants marked by black dots in (c). (h) The Hovm\"{o}ller plot of $\theta$ corresponds to the TPR solution shown in (c) with $x = \pi/k_c$ and $z = 0.5$.}
\label{AR2_Fig2}
\end{figure}

For any angle $\phi$, convection begins either as SPR or SOR$^+$, corresponding to smaller and larger values of $\mathrm{Q}$, respectively. Increasing $r$ reveals the transition from SPR and SOR$^+$ to chaos through three sequences of solutions: two for SPR and one for SOR$^+$. For a lower magnetic field, $\mathrm{Q}=10$, we observe transient chaos followed by persistent chaos as $r$ is increased. This scenario is shown in the figure~\ref{AR2_Fig1}. Figure~\ref{AR2_Fig1}(a) shows the time evolution of the important modes corresponding to the SPR flow patterns for $r=3$ where $W_{011}\neq 0$, and $W_{111}=W_{-111} = 0.$ For higher  $r$ we observe transient chaos as shown for $r = 5.62$ in the figure~\ref{AR2_Fig1}(b) before settling to SPR in the long run. The time interval of the initial chaotic solution gradually increases with $r$ and eventually, the system exhibits a chaotic solution as shown in the figure~\ref{AR2_Fig1}(c) which is manifested as chaotic traveling parallel rolls (CTPR) flow patterns. With a slight increase in the strength of magnetic field (e.g., $\mathrm{Q}=30$), SPR appears again at the onset of convection while increasing r leads to different set of flow patterns.  At sufficiently high $r$, the system eventually becomes chaotic exhibiting CTPR flow patterns through intermediate PPR and traveling parallel rolls (TPR) flow regimes. The sequence of the related time evolution of the important Fourier modes corresponding to the qualitatively different flow regimes starting from the  SPR is shown in the figures~\ref{AR2_Fig2}(a) - (d). The isotherms showing the traveling rolls flow patterns corresponding to the marked instants on the figure~\ref{AR2_Fig2}(c) are shown in the figures~\ref{AR2_Fig2}(e)-(g). To confirm the traveling nature of the flow pattern we also construct a Hovm\"oller plot shown in the figure~\ref{AR2_Fig2}(h). The oblique orientation of the plot confirms the traveling nature of the flow pattern. For higher magnetic fields, convection begins with SOR$^+$. As r increases, the flow progresses through SPR, PPR, TPR, and finally CTPR, similar to the bifurcation scenario for $\mathrm{Q}=30$ (Figs. \ref{AR2_Fig2}(a) - (d)).

\begin{figure}[h]
\centering
\includegraphics[scale = .35]{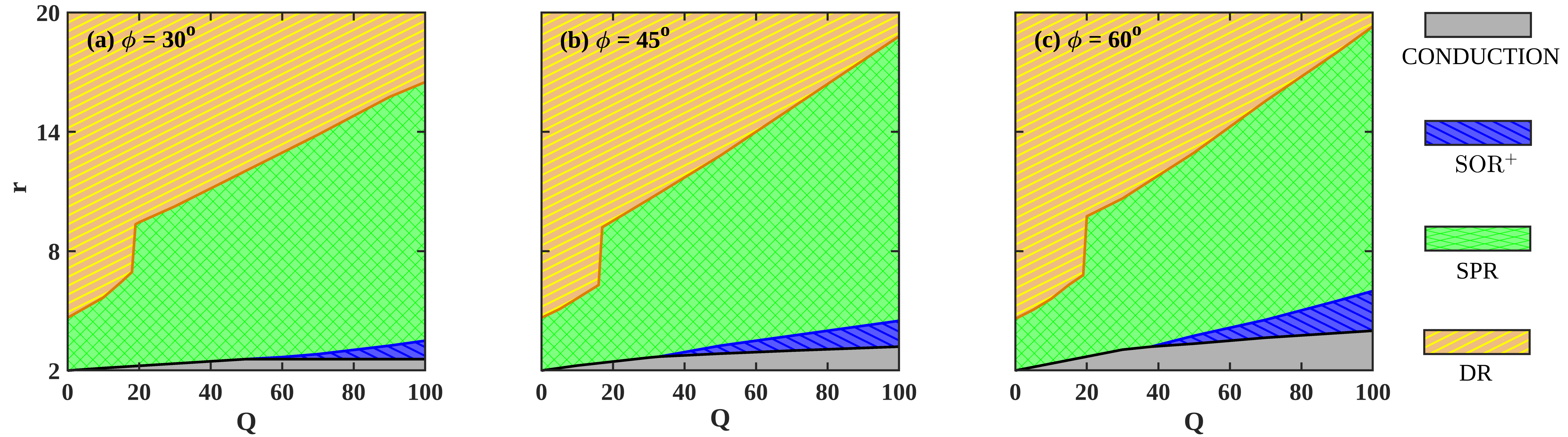}
\caption{(a) - (c) Two-parameter diagrams showing different flow regimes on the $\mathrm{Q} - r$ plane for $\mathrm{Pr} = 0.1$, $\Gamma = 2$ and three values of $\phi$.}
\label{AR2_Fig3}
\end{figure}

Finally, for a detailed understanding of the flow regimes in the considered parameter space, we prepare three two-parameter diagrams from the DNS data, one each for $\phi = 30^\circ$, $45^\circ$, and $60^\circ$. The diagrams are shown in the figure~\ref{AR2_Fig3}. Different flow regimes are clearly demarcated and represented with different colors. The diagrams reveal three types of flow regimes, namely, SOR$^+$, SPR, and dynamic regime. As the strength of the magnetic field increases, the stationary regime increases, while the dynamic regime decreases. Thus, the magnetic field is found to play a key role in stabilizing the flow. 

\section{Conclusions}
In summary, we have numerically investigated the effect of external horizontal oblique magnetic field on the convection rolls occurring near the onset of convection in low Prandtl number electrically conducting fluids considering the classical Rayleigh-B\'{e}nard convection system. A wide region of the parameter space given by $0\leq \mathrm{Q} \leq 1000$ and $1 \leq r \leq 20$ has been explored by performing detailed three dimensional direct numerical simulations for three horizontal aspect ratios. 

The investigation reveals a very rich bifurcation scenario near the onset involving three types of stationary flow patterns, namely, SPR, SOR$^+$, and SOR$^-$. In the absence of an external oblique magnetic field, convection starts in the form of SPR. The Lorentz force arising out of the transverse component of the oblique magnetic field suppresses convection in the form of the rolls, while the longitudinal component does not influence it.                                                                                                                                      However, the secondary and higher-order instabilities occurring for the larger Rayleigh numbers are greatly affected by both components. 

For the square aspect ratio cell ($\Gamma = 1$), the convection restarts in the form of SPR for all $\phi \in [0^\circ, 90^\circ]$ when the external magnetic field is weak. On the other hand, for the stronger magnetic field, the convection restarts in the form of SPR for lower values of $\phi$, SOR$^+$ for larger values of $\phi ~(< 90^\circ)$ and  SOR$^-$ for $\phi = 90^\circ$. Beyond the onset as $r$ is increased in each of the cases, the oscillatory instability of the steady rolls is observed. Interestingly, the onset of the OI of the steady rolls is found to scale as $\mathrm{Q}^\alpha$ with two distinct exponents, one each for weaker and stronger magnetic fields respectively. Moreover, two different routes to chaos are observed. For, weaker and stronger magnetic fields, period doubling and quasiperiodic routes to chaos respectively are observed. This observation is consistent with the experimental observation~\cite{fauve_prl:1984}.  It is also observed that for a given set of values of $\mathrm{Q}$ and $\mathrm{Pr}$, the heat transfer is inhibited with the increase of $\phi$. For a very strong external magnetic field, a new type of stationary oblique flow pattern governed by the most energetic mode $W_{-211}$ is observed.

As the aspect ratio of the computational box is made rectangular, interesting new dynamics are observed. We have considered two rectangular cells given by $\Gamma = \frac{1}{2}$ and $\Gamma = 2$. In both cases, a transition to traveling wave flow regimes is observed with the increase of $r$, for fixed $\mathrm{Q}$, $\mathrm{Pr}$ and $\phi$. Additionally, we also observe the existence of transient chaos followed by chaos for $\Gamma = \frac{1}{2}$. A detailed understanding of the flow regimes is developed in a wide region of the parameter space. We expect that the present study will be helpful in controlling a convective flow of electrically conducting fluids by applying an external magnetic field. \\

\noindent{{ACKNOWLEDGMENT}}\\
S.S. acknowledges the support from the University Grants Commission, India (Award No. 191620126754).

\end{document}